\documentclass[journal,onecolumn,romanappendices]{IEEEtran}

\usepackage{url}
\usepackage{ifthen}
\usepackage[cmex10]{amsmath} 
\usepackage{longtable}
\usepackage{makecell}
\usepackage{algorithm,algorithmic}
\usepackage{color}
\usepackage[normalem]{ulem}
                             
\usepackage{amssymb,amsthm}
\usepackage{mathtools,bbm}  
\usepackage{xifthen,xparse}
\usepackage{hyperref}  
\usepackage{float}
\usepackage{hhline}
\usepackage{caption}

\usepackage{stmaryrd}

\usepackage{textcomp}

\allowdisplaybreaks

\newtheorem{theorem}{Theorem}
\newtheorem{result}{Result}

\newtheorem{remark}[theorem]{Remark}
\newtheorem{lemma}[theorem]{Lemma}
\newtheorem{definition}[theorem]{Definition}
\newtheorem{example}{Example}

\newenvironment{example*}
  {\addtocounter{example}{-1}\example}
  {\endexample}

\newcommand{\tr}[1]{\text{Tr}\left[ #1 \right]}
\newcommand{\norm}[1]{\left\| #1 \right\|}	
\newcommand{\dket}[1]{\left\lvert #1 \right\rangle}
\newcommand{\dbra}[1]{\left\langle #1 \right\rvert}
\NewDocumentCommand\dketbra{+m+g}{%
  \IfNoValueTF{#2}
    {\left\lvert #1 \right\rangle \left\langle #1 \right\vert}
  {\left\lvert #1 \right\rangle \left\langle #2 \right\rvert}%
}
\NewDocumentCommand\dbraket{+m+g}{%
  \IfNoValueTF{#2}
    {\left\langle #1 \vert #1 \right\rangle}
  {\left\langle #1 \vert #2 \right\rangle}%
}

\global\long\def\topbotatom#1{\hbox{\hbox to0pt{\ensuremath{#1{\scriptstyle \bot}}\hss}\ensuremath{#1{\scriptstyle \top}}}}
 \global\long\def\topbot{{\mathord{\mathchoice{\displaystyle \topbotatom{}}{\textstyle \topbotatom{}}{\scriptstyle \topbotatom{}}{\scriptscriptstyle \topbotatom{}}}}}

\global\long\def\rank{\mathrm{rank}}

\newcommand{\MCC}{\mathcal{C}}
\newcommand{\MCCd}{\mathcal{C}^{\perp}}

\newcommand{\vecnot}[1]{\underline{#1}}

\newcommand{\pr}[1]{\mathbb{P}\left( #1 \right)}

\newcommand{\psc}{\text{PSC}}
\newcommand{\bsc}{\text{BSC}}
\newcommand{\Ag}{A}
\newcommand{\Bg}{B}

\renewcommand{\triangleq}{\coloneqq}

\DeclareMathOperator*{\argmax}{arg\,max}

\newif\ifnotes
\notestrue

\begin{document}

\title{A Semiclassical Proof of Duality Between\\ the Classical BSC and the Quantum PSC}

 \author{%
   \IEEEauthorblockN{Narayanan Rengaswamy
                     and
                     Henry D. Pfister}%
   \thanks{
           N. Rengaswamy is with the 
           Department of Electrical and Computer Engineering,
           University of Arizona, Tucson, Arizona 85721, USA.
           H. D. Pfister is with the
           Department of Electrical and Computer Engineering,
           Duke University,
           Durham, North Carolina 27708, USA.
           Most of this work was conducted when N. Rengaswamy was with the Department of Electrical and Computer Engineering,
           Duke University,
           Durham, North Carolina 27708, USA.
           Email: narayananr@arizona.edu, henry.pfister@duke.edu}%
  }



\maketitle

\begin{abstract}
In 2018, Renes [IEEE Trans. Inf. Theory, vol. 64, no. 1, pp. 577-592 (2018)] developed a general theory of channel duality for classical-input quantum-output (CQ) channels.
That result showed that a number of well-known duality results for linear codes on the binary erasure channel could be extended to general classical channels at the expense of using dual problems which are intrinsically quantum mechanical.
One special case of this duality is a connection between coding for error correction (resp. wire-tap secrecy) on the quantum pure-state channel (PSC) and coding for wire-tap secrecy (resp. error correction) on the classical binary symmetric channel (BSC).
While this result has important implications for classical coding, the machinery behind the general duality result is rather challenging for researchers without a strong background in quantum information theory.
In this work, we leverage prior results for linear codes on PSCs to give an alternate derivation of the aforementioned special case by computing closed-form expressions for the performance metrics.
The noted prior results include optimality of the square-root measurement (SRM) for linear codes on the PSC and the Fourier duality of linear codes.
We also show that the SRM forms a suboptimal measurement for channel coding on the BSC (when interpreted as a CQ problem) and secret communications on the PSC.
Our proofs only require linear algebra and basic group theory, though we use the quantum Dirac notation for convenience.
\end{abstract}

\section{Introduction}
\label{sec:intro}

In the mathematical sciences, duality is a powerful concept that connects two problems such that the solution of one determines the solution of the other.
In coding theory, an $[n,k]$ binary linear code $\mathcal{C} \subseteq \mathbb{F}_2^n$ is a $k$-dimensional subspace of the vector space of length-$n$ binary vectors.
In this case, its dual code $\mathcal{C}^\perp \subseteq \mathbb{F}_2^n$ is the $(n-k)$-dimensional subspace that is orthogonal to $\mathcal{C}$ under the standard (binary) dot product.
An early and important implication of this duality is that the weight enumerator (WE) of a linear code can be computed from the WE of the dual code using the MacWilliams identity~\cite{MacWilliams-bell63}.

For classical channels, the notion of a dual channel did not arise until after the rediscovery of low-density parity-check (LDPC) codes, and then it was understood only for the erasure channel.
Let BEC($\epsilon$) denote the binary erasure channel with erasure probability $\epsilon$.
It was shown in \cite{Ashikhmin-it04} that the extrinsic information transfer (EXIT) function of a code on the BEC($\epsilon$) is closely related to the EXIT function of the dual code on the BEC($1-\epsilon$).
This and other symmetries in the decoding analysis of LDPC codes led some researchers to treat the BEC($1-\epsilon$) as the dual channel of the BEC($\epsilon$)~\cite{Chung-2000,Martinian-allerton03,Ashikhmin-it04,Pfister-it07,Thangaraj-it07,Obata-isit13}.

For more than 10 years, it remained an open question whether this notion of channel duality could be extended beyond the erasure case.
In 2018, Renes~\cite{Renes-it18} provided such a definition by showing that the dual channel of a binary memoryless channel can be defined in terms of a \emph{classical-input quantum-output} (CQ) channel.
In particular, Renes developed a general theory of CQ channel duality where a channel $W$ and its dual $W^\perp$, both CQ channels with $d$ input symbols, satisfy $H(W)+H^\perp (W^\perp) = \log d$ for primal and dual entropies $H$ and $H^\perp$.
We do not discuss this general duality further but provide an alternative operational perspective for some special cases.
For example, if we let $W=\text{BEC}(\epsilon)$ then \cite{Renes-it18} shows that $W^\perp = \text{BEC}(1-\epsilon)$ and recovers some previously known results for the BEC.
Additionally, if $W = \psc(\theta)$ is the CQ binary \emph{pure-state channel (PSC)} with parameter $\theta$~\cite{Renes-njp17}, then $W^\perp = \bsc(p)$ is the classical binary symmetric channel (BSC) with parameter $p = (1 - \cos\theta)/2$.
Note that any classical channel can be  treated as a CQ channel by defining the outputs to be diagonal in the standard basis, i.e., $0 \equiv \dket{0}, 1 \equiv \dket{1}$.
Thus, just as the complex numbers now play an important role in our understanding of the real numbers, this shows that CQ channels are an inherent part of the theory of classical channels (e.g., see also~\cite{Dalai-it13}).
Moreover, Renes also extends extrinsic information transfer (EXIT) function duality for linear codes on the BEC to general CQ channels in~\cite{Renes-it18}.

This paper considers dualities between communication problems, linear codes, and performance metrics.
To start, let us discuss the single channel setup.
For a CQ channel $W$ with input $X \in \mathcal{X}$ and output $B$, this is based on the quantum conditional entropies $H_{\min}(X|B)$ and $H_{\max}(X|B)$ which will be defined shortly.
Consider a classical system $X$ that is coupled to quantum system $B$.
For any measurement $\mathcal{M}$ of the quantum system $B$, let $P_{X|M}$ denote the posterior of $X$ given the measurement outcome $M$ and let the maximum successful guessing probability~\cite{Koenig-it09,Wilde-2013} be given by
\begin{align}
P(W) \triangleq \max_\mathcal{M} \mathbb{E} \Big[\max_x P_{X|M}(x|M)\Big],
\end{align}
where the maximum is over all quantum measurements of $B$.
Then, $H_{\min}(X|B) \triangleq - \log_2 P(W)$.
Similarly, for secret communication where an eavesdropper observes $X$ through $W$, one can measure of information leakage is
\begin{align}
Q(W) \triangleq \min_\mathcal{M} \mathbb{E}\big[\mathcal{B}(P_X,P_{X|M})\big]^2,
\end{align}
where $\mathcal{B}(p,q) \triangleq \sum_{x\in \mathcal{X}} \sqrt{p(x) q(x)}$ is the Bhattacharyya coefficient\footnote{The Bhattacharyya distance between two distributions is typically defined by $-\log \mathcal{B}(p,q)$ so that its value equals 0 if and only if $p=q$.} between the two pmfs~\cite[p.~221]{Koenig-it09,Wilde-2013}.
Then, $H_{\max}(X|B) \triangleq \log_2| \mathcal{X}| + \log_2 Q(W)$. 
Moreover, the entropies $H_{\min}$ and $H_{\max}$ are dual in the sense that $H_{\min}(W)+H_{\max} (W^\perp) = \log |\mathcal{X}|$~\cite{Renes-it18}.
An equivalent description of this entropic duality is that $P(W) = Q(W^\perp)$.

In \cite{Renes-it18}, the above is also extended to the case where linear codes are used for both problems.
This is based on defining a CQ superchannel whose input symbols are associated with codewords and whose output is an observation of the codeword through a CQ channel.
In this paper, we give an alternate derivation of that result
for the PSC-BSC pair by directly calculating closed-form expressions for (i) the block error rate and (ii) the Bhattacharyya distance between the posterior distribution of the secret message and the uniform distribution.
Our approach also establishes some results from~\cite{Renes-it18} for Von Neumann entropy including the above coding-secrecy result and a new duality result for generalized EXIT (GEXIT) functions.
While the approach in \cite{Renes-it18} uses some sophisticated quantum techniques, our exposition relies on direct calculation and targets an audience of classical information and coding theorists.


\subsection{Summary of Results}

\begin{result}[Theorem~\ref{thm:Pe_PSC}]
Consider using an $[n,k]$ binary linear code $\MCC$ with generator matrix $G$ to transmit $k$ bits over the $\psc(\theta)$.
Then, the probability of block error for the optimal detector is given by
\begin{equation}
P_e = 1 - \mathcal{B}\left( P_{S'|X}(\cdot|x), \nu (\cdot) \right)^2,
\end{equation}
where $\nu (s')= 2^{-k}$ is the uniform distribution on $\{0,1\}^k$ and $P_{S'|X}(s'|x)$ is the conditional probability of the secret message $S'$ given the observation $X$ for Wyner's wire-tap coding scheme over $\bsc\left( \frac{1-\cos\theta}{2} \right)$, with $S'$ indexing the cosets of $\MCCd$ (see Sections~\ref{sec:cc_sec_duality_bec} and~\ref{sec:cc_psc}).
This follows from writing $P_e$ in terms of $r(u) = (\cos\theta )^{w_H (uG)}$, where $w_H(\cdot)$ is the Hamming weight, and observing that
\begin{equation}
P_{S'|X}(s'|x) = \frac{1}{2^{k/2}} \hat{r}(\pi_x (s') ),
\end{equation}
where $\hat{r}(t) = 2^{-k/2} \sum_{u\in\{0,1\}^k} (-1)^{tu^T} r(u)$ is the scaled Fourier transform of $r(u)$ and $\pi_x$ is a permutation on $\{0,1\}^k$ for each $x \in \{0,1\}^n$ (see Section~\ref{sec:wyner_bsc}).
Notice that $P_e$ does not depend on $x$ because $\pi_x$ is a permutation and $\nu(\cdot)$ is constant.

\end{result}

This is an example of the fact (from \cite{Renes-it18}) that $H_{\min}(W)+H_{\max} (W^\perp) = \log |\mathcal{X}|$ because 
$$1 - P_e = P(W) = 2^{-H_{\min}(W)} = \frac{1}{|\mathcal{X}|} 2^{H_{\max} (W^\perp)} = Q(W^\perp) = \mathcal{B}\left( P_{S'|X}(\cdot|x), \nu (\cdot) \right)^2,$$ 
where $W = \psc(\theta), W^\perp = \bsc\left( \frac{1-\cos\theta}{2} \right),$ and $|\mathcal{X}| = 2^k$.
We refer to our derivation as semiclassical because it uses a minimal amount of quantum theory and the dual problem is entirely classical.
In particular, the proof is explicit and makes use of the group-theoretic results for the SRM
by Eldar and Forney~\cite{Eldar-it00} and Fourier duality of linear codes~\cite{Hartmann-it76,Forney-arxiv11,Pfister-fg_duality} (see Appendix~\ref{sec:fg_duality}).

Next, we swap the PSC and BSC, i.e., consider channel coding over the BSC and secret communications over the PSC.
To do this, we embed the classical BSC problem in a CQ setup.
Given any vector $v \in \{0,1\}^n$, let $v^*$ denote a minimum-weight vector in the coset $v \oplus \MCCd$ (with ties broken arbitrarily).

\begin{result}[Theorems~\ref{thm:BSC_MAP} and~\ref{thm:BSC_SRM}]
The maximum-a-posteriori (MAP) decoder for channel coding on the $\bsc(p)$ with the $[n,n-k]$ binary linear code $\MCCd$ can be implemented using the projective measurement defined by
\begin{align}
\left\{ \Pi_z \coloneqq \sum_{v \in \MCC^{\topbot}} \dketbra{c_z \oplus v^*} ; \ z \in \mathbb{Z}_2^{n-k} \right\}.
\end{align}
Here, $c_z$ denotes the codeword in $\MCCd$ for the message $z$, and $\MCC^{\topbot}$ refers to a code that is complementary to the code $\MCCd$ (see Section~\ref{sec:linear_codes}).
In other words, this measurement achieves the well-known optimal probability of success given by
\begin{align}
\mathbb{P}\left[ \text{MAP success for $\MCC^{\perp}$ on BSC$(p)$} \right] = \sum_{v \in \MCC^\topbot} \max_{ u \in (v \oplus \MCC^{\perp}) } p^{w_H(u)} (1-p)^{n - w_H(u)} = \sum_{v \in \MCC^\topbot} p^{w_H(v^*)} (1-p)^{n - w_H(v^*)},
\end{align}
where $w_H(u)$ refers to the Hamming weight of $u$.

Furthermore, the SRM is inferior to this MAP measurement (decoder) and its success probability is given by 
\begin{align}
\!\!\mathbb{P} \left[ \text{SRM success for $\MCC^{\perp}$ on BSC$(p)$} \right] 
  & = \sum_{v \in \MCC^\topbot} \dfrac{\sum_{u \in (v \oplus \MCC^{\perp})} \left( p^{w_H(u)} (1-p)^{n - w_H(u)} \right)^2}{ \sum_{u \in (v \oplus \MCC^{\perp})} p^{w_H(u)} (1-p)^{n - w_H(u)} }.
\end{align}
\end{result}

\begin{result}[Theorems~\ref{thm:psc_secrecy_fidelity} and~\ref{thm:psc_secrecy_srm_suboptimal}]
Consider secret communication over the $\psc(\theta)$ using the cosets of $\MCC$, 
which are indexed by the complementary code $\MCC^\top$ (see Section~\ref{sec:linear_codes}).
Let us define $\beta(v) \triangleq 2^{n-k} p_{v^*}/q = 2^{n-k} p^{w_H(v^*)} (1 - p)^{n - w_H(v^*)}/q$, where $q \triangleq \sum_{v \in \MCC^{\topbot}} p_{v^*}$.
Then, the optimal choice for $\sigma_{B^n}$ (for which the fidelity in $H_{\max} (W^\perp)$~\eqref{eq:secrecy_psc_fidelity} equals the MAP success rate $q$ for $\MCCd$ on the BSC($p$)) is 
\begin{equation}
\sigma_{B^n} = \frac{1}{2^{n-k}} \sum_{v \in \MCC^{\topbot}} \beta(v) \dketbra{v^*}.
\end{equation}
Furthermore, using the identity connecting the quantum fidelity and the classical Bhattacharyya distance~\cite[Chapter 9]{Wilde-2013}, we see that the SRM does not induce the optimal fidelity.
\end{result}

Once again, this result is a specific example (with explicit details) of the fact that $H_{\min}(W) + H_{\max} (W^\perp) = \log |\mathcal{X}|$~\cite{Renes-it18} for $W = \bsc\left( \frac{1-\cos\theta}{2} \right)$, $W^\perp = \psc(\theta)$, and $|\mathcal{X}| = 2^{n-k}$.

The SRM, which is also called the pretty-good measurement (PGM), is an important and useful measurement for many quantum tasks.
We have used the optimality of the SRM to derive the optimal block-error probability above for channel coding over the PSC.
However, it is interesting to see that it is suboptimal for the purely classical problem of channel coding over the BSC.
This is perhaps because, when the transmitted codeword is fixed, the output of the BSC still has some randomness, whereas the PSC output is deterministic but cannot be directly observed like a classical vector (see Section~\ref{sec:psc}).

\begin{remark}
This SRM analysis has also been used recently for channel coding over the PSC to verify that belief propagation with quantum messages (BPQM) is quantum-optimal with respect to the block success probability for a $5$-bit code~\cite{Rengaswamy-arxiv20}.
Since BPQM produces a structured receiver circuit,
this connection enables one to design practical receivers for optical communications over pure-loss bosonic channels.
\end{remark}

The paper is organized as follows.
Section~\ref{sec:bec_duality} discusses the duality between channel coding and secret communications in the context of the binary erasure channel.
This establishes some elements of the general duality result, between the PSC and BSC, in a purely classical setting.
Section~\ref{sec:background} introduces necessary background on quantum concepts so that classical information and coding theorists can follow this PSC-BSC case using standard linear algebra and a modicum of group theory.
Moreover, Section~\ref{sec:linear_codes} discusses an alternative perspective on classical binary linear codes introduced by Renes in~\cite{Renes-it18}.
Then, Section~\ref{sec:duality_coding_secrecy} discusses channel coding and secret communications over the PSC and BSC in detail.
Finally, Section~\ref{sec:conclusion} concludes the paper and discusses potential future work.
Throughout the paper, we make remarks that might provide additional insights.

\section{Duality Between Channel Coding and Secret Communication for the BEC}
\label{sec:bec_duality}

\subsection{Duality for Channel Coding on the BEC}

Let $\mathcal{C}$ be an $(n,k)$ binary linear code with generator matrix $G$ and parity-check matrix $H$. Assume that a random codeword $X$ is chosen uniformly and transmitted through a BEC with output $Y\in\{0,1,?\}^{n}$.
For an output realization $y$, let $\mathcal{E} \triangleq\left\{ i\in[n]\,|\,y_{i}=?\right\} $ be the set of indices where an erasure occurs.
It turns out that many duality statements are more natural for a deterministic length-$n$ BEC that erases all bits whose indices are in $\mathcal{E}$.
We refer to this channel as BEC$(\mathcal{E})$, and the erasure patten $\mathcal{E}$ is fixed.
Its dual channel, which correctly transmits only the bits with indices in $\mathcal{E}$, is denoted by BEC$(\mathcal{E}^c)$.

For a set $\mathcal{E}=(e_{1},e_{2},\ldots,e_{|\mathcal{E}|})\subset \mathbb{Z}$ with $1\leq e_{1}<e_{2}<\cdots<e_{|\mathcal{E}|} \leq n$ and an $m\times n$ matrix $G=(g_{1},g_{2},\ldots,g_{n})$ whose $i$-th column is $g_i$, we let $G_{\mathcal{E}}=(g_{e_{1}},g_{e_{2}},\ldots,g_{e_{|\mathcal{E}|}})$ be an $m \times |\mathcal{E}|$ matrix.
We also use this notation for row vectors with $m=1$. 
Let $V=\left\{ z\in \mathcal{C}\,|\,z_{\mathcal{E}^{c}}=y_{\mathcal{E}^{c}}\right\} $ be the set of codewords that are compatible with the observations.
Then, the posterior distribution of $X$ given $Y$ is 
$P_{X|Y}(x|y)= 1/|V|$ if $x\in V$ and 0 otherwise.

Since $\mathcal{C}$ is linear, the set $V$ is the affine subspace of $x\in\{0,1\}^{n}$ satisfying 
$H_{\mathcal{E}}x_{\mathcal{E}}^{T}=H_{\mathcal{E}^{c}}x_{\mathcal{E}^{c}}^{T}$ because $x_{\mathcal{E}^c}$ is known at the decoder.
Thus, dimension of the solution space is given by $|\mathcal{E}|-\rank(H_{\mathcal{E}})$. Similarly, the affine subspace of input vectors $u\in\{0,1\}^{k}$ compatible with $y$ is defined by
$uG_{\mathcal{E}^{c}}=x_{\mathcal{E}^{c}}$
and dimension of the solution space is $k-\rank(G_{\mathcal{E}^{c}})$. Of course, the two spaces must have the same dimension and this implies that
$k-\rank(G_{\mathcal{E}^{c}})=|\mathcal{E}|-\rank(H_{\mathcal{E}})$.
Thus, we find that
\begin{equation} \label{eq:bec_coding_dual}
H(X|Y) = H(X|X_{\mathcal{E}^c}) =|\mathcal{E}|-\rank(H_{\mathcal{E}}) =k-\rank(G_{\mathcal{E}^{c}}).
\end{equation}

Let $X' \in \mathcal{C}^\perp$ be a uniform random dual codeword and $Y'=X_{\mathcal{E}}'$ be its observation through the dual channel, BEC$(\mathcal{E}^c)$.
Then, the first equality in~\eqref{eq:bec_coding_dual} shows that the entropy of $X'$ given the dual observation is
\begin{equation}
H(X'|Y') = H (X'|X'_{\mathcal{E}}) = |\mathcal{E}^c| - \rank(H_{\mathcal{E}^c}^{\perp}) = |\mathcal{E}^c| - \rank(G_{\mathcal{E}^c})
\end{equation}
because $H^\perp = G$.
Using~\eqref{eq:bec_coding_dual} to substitute for $\rank(G_{\mathcal{E}^c})$ gives
\begin{equation}
H (X'|X'_{\mathcal{E}}) = H(X|X_{\mathcal{E}^c}) + |\mathcal{E}^c| - k. \label{eq:entropy_duality}
\end{equation}
As we will see in~\eqref{eq:vne_coding_duality}, if we rewrite \eqref{eq:entropy_duality} as $H (X'|Y') = H(X|Y) + \text{``dual entropy"} - k$, then it generalizes to the case where $Y$ is a PSC$(\theta)$ observation of $X$ and $Y'$ is a BSC$(\frac{1-\cos\theta}{2})$ observation of $X$.

\subsection{Duality Between Channel Coding and Secrecy on the BEC}
\label{sec:cc_sec_duality_bec}

In 1975, Wyner introduced the wire-tap channel and proposed encoding secret messages into cosets of a group code \cite{Wyner-it75}. Encoding proceeds by using the secret message to choose a coset and then encoding to a uniform random element from that coset. In this section, we will see that there is a duality between the information loss of channel coding using $\mathcal{C}$ and the information leakage of Wyner's coset coding using $\mathcal{C}^{\perp}$.

First, we will consider 
the standard channel coding problem for $\mathcal{C}$ and Wyner's wire-tap coding using cosets of $\mathcal{C}$. The coding problem transmits the codeword $x\in\{0,1\}^{n}$ as determined by the information $u\in\{0,1\}^{k}$ and coset selector $s\in\{0,1\}^{n-k}$ using the definitions
\begin{align}
\Ag = 
\begin{bmatrix}
G\\
F
\end{bmatrix}, \quad x= [u \;\, s]\, \Ag =[u \;\, s]\begin{bmatrix}G\\
F
\end{bmatrix}=uG+sF.
\end{align}
In this setup, $\text{rowspace}(F)$ is a linear complement of $\mathcal{C}=\text{rowspace}(G)$ and $\Ag$ is full rank. 
We note that the definition of $A$, and the use of $A^{-1}$ below, is motivated by~\cite{Renes-it18}.

In the channel coding problem, one assumes that the receiver knows the coset vector $sF$. In contrast, the wire-tap coding problem assumes $u$ is unknown and tries to decode the secret message $s$. To make this stochastic, we let $X$ be a uniform random vector over $\{0,1\}^{n}$ and define the random vectors $U\in\{0,1\}^{k}$ and $S\in\{0,1\}^{n-k}$ via 
$
[U \; S]=X \Ag^{-1}.
$
We note that choosing the uniform distribution for $X$ is equivalent to using uniform distributions for $U$ and $S$.

Next, we let $Y=X_{\mathcal{E}^c}$ be the BEC$(\mathcal{E})$ observation of $X$.
Then, we can write
\begin{align}
n & =I(U,S;X) \nonumber \\
 & =I(U,S;X_{\mathcal{E}^{c}},X_{\mathcal{E}}) \nonumber \\
 & =I(U,S;X_{\mathcal{E}^{c}})+I(U,S;X_{\mathcal{E}}|X_{\mathcal{E}^{c}}) \nonumber \\
 & =I(S;X_{\mathcal{E}^{c}})+I(U;X_{\mathcal{E}^{c}}|S)+I(U,S;X_{\mathcal{E}}|X_{\mathcal{E}^{c}}) \nonumber \\
 & =I(S;X_{\mathcal{E}^{c}})+I(U;X_{\mathcal{E}^{c}}|S)+\left|\mathcal{E}\right| \nonumber \\
 & =I(S;X_{\mathcal{E}^{c}})+k-H(U|X_{\mathcal{E}^{c}},S)+\left|\mathcal{E}\right|.
\end{align}
The final equation relates the the number of erasures $\left|\mathcal{E}\right|$ to the information leakage $I(S;Y)=I(S;X_{\mathcal{E}^{c}})$ of the secrecy problem and to the message uncertainty $H(U|Y,S)=H(U|X_{\mathcal{E}^{c}},S)$ of the coding problem. Using $I(S;X_{\mathcal{E}^{c}})=n-k-H(S|X_{\mathcal{E}^{c}})$, we also see that
\begin{align}
H(S|X_{\mathcal{E}^{c}})+H(U|X_{\mathcal{E}^{c}},S)=\left|\mathcal{E}\right|.
\end{align}
From this formula, it follows that perfect secrecy is achieved for an erasure pattern with $\left|\mathcal{E}\right|=k$ if and only if that erasure pattern is correctable in the coding problem (i.e., $H(U|S,X_{\mathcal{E}^{c}})=0$). Closely related statements of this type have appeared in a few prior works~\cite{Wyner-it75,Martinian-allerton03,Thangaraj-it07}. Thus, good codes for coding (i.e., with small $H(U|X_{\mathcal{E}^{c}},S)$) are good for secrecy (i.e., have large $H(S|X_{\mathcal{E}^{c}})$). 

Since $X$ is uniform, we can use a similar trick to interpret it as the codeword for coding/secrecy problem for the dual code $\mathcal{C}^{\perp}$ on the dual channel, where the erased positions are in $\mathcal{E}^c$.
Note that \vspace{-0.5mm}
\begin{align}
x \Ag^{-1}=[u \;\, s]\Ag \Ag^{-1}=[u \;\, s] \vspace*{-0.5mm}
\end{align}
implies that the last $k$ columns of $\Ag^{-1}$ give the transpose of a parity-check matrix $H$ for $\mathcal{C}$. Also, the first $n-k$ columns of $\Ag^{-1}$ give a right inverse for $G$ which we denote by $E^{T}$ (i.e,. $GE^{T}=I)$. Thus, we can define $\Bg = (\Ag^{-1})^{T}$,
\begin{align}
\Bg = 
\begin{bmatrix}
E\\
H
\end{bmatrix}, \quad
x=[s' \; u']\, \Bg = [s' \; u']\begin{bmatrix}E\\
H
\end{bmatrix}=s'E+u'H,
\end{align}
and view any $x\in\{0,1\}^{n}$ as the sum of the dual codeword $u'H$ and the coset vector $s'E$. We also define the random vectors $U'\in\{0,1\}^{n-k}$ and $S'\in\{0,1\}^{k}$ via 
$
[S' \; U']=X \Bg^{-1}.
$

The secrecy and coding problems for the dual code on the dual channel can be related by swapping $\mathcal{E}$ and $\mathcal{E}^c$.
This gives
\begin{align}
n &=I(U',S';X) \nonumber \\
 & =I(U',S';X_{\mathcal{E}},X_{\mathcal{E}^{c}}) \nonumber \\
 & =I(U',S';X_{\mathcal{E}})+I(U',S';X_{\mathcal{E}^{c}}|X_{\mathcal{E}}) \nonumber \\
 & =I(S';X_{\mathcal{E}})+I(U';X_{\mathcal{E}}|S')+I(U',S';X_{\mathcal{E}^{c}}|X_{\mathcal{E}}) \nonumber \\
 & =I(S';X_{\mathcal{E}})+I(U';X_{\mathcal{E}}|S')+(n-\left|\mathcal{E}\right|) \nonumber \\
 & =I(S';X_{\mathcal{E}})+(n-k)-H(U'|X_{\mathcal{E}},S')+(n-\left|\mathcal{E}\right|). \label{eq:bec_mutual_deriv}
\end{align}
The result in~\eqref{eq:bec_mutual_deriv} relates the number of erasures $|\mathcal{E}|$, the information leakage $I(S';X_{\mathcal{E}})$ of the dual-code secrecy problem on the dual channel, and the message uncertainty $H(U'|X_{\mathcal{E}},S')$ for the coding problem using the dual code and dual channel.
Now, we can use~\eqref{eq:entropy_duality} to substitute  the uncertainty for primal coding on the primal channel, $H(U|X_{\mathcal{E}^{c}},S)$, for $H(U'|X_{\mathcal{E}},S')$ and a few other terms.
This gives
\begin{align}
H(U'|X_{\mathcal{E}},S')=H(U|X_{\mathcal{E}^{c}},S)+n-k-\left|\mathcal{E}\right|.
\end{align}
For the dual code and channel, combining with~\eqref{eq:bec_mutual_deriv} allows us to write the information leakage of the secrecy problem as
\begin{align}
H (S'|X_{\mathcal{E}}) & = k - I(S';X_{\mathcal{E}}) \nonumber \\
 & = k - \left( k + H(U'|X_{\mathcal{E}},S')- n + \left|\mathcal{E}\right| \right) \nonumber \\
 & = k - H(U|X_{\mathcal{E}^{c}},S) \\
 & = k - H(U|X_{\mathcal{E}^{c}},S=0),\label{eq:dualsecrecy_vs_primalcoding}
\end{align}
where the last equality holds due to channel symmetry.
As we will see in~\eqref{eq:neumann_duality}, if we rewrite \eqref{eq:dualsecrecy_vs_primalcoding} as $H (S'|Y') = k-H(U|Y,S=0)$, then it also holds for the more general case where $Y$ is a PSC$(\theta)$ observation of $X$ and $Y'$ is a BSC$(\frac{1-\cos\theta}{2})$ observation of $X$.

\subsubsection*{Block Error Rate and Bhattacharyya Distance}

If the uncertainty in $U$ has dimension $d=H(U|X_{\mathcal{E}^{c}},S=0)$, then the probability of correctly guessing the primal codeword is $2^{-d}$. 
If the posterior of $S'$ given $X_{\mathcal{E}}$ is uniform over an affine subspace of dimension $f=H (S'|X_{\mathcal{E}})$, then \eqref{eq:dualsecrecy_vs_primalcoding} implies $f=k-d$ and the Bhattacharyya coefficient between this posterior and the uniform distribution is \vspace*{-0.5mm}
\[
\sum_{s\in\{0,1\}^{f}}\sqrt{p_{s}2^{-k}}=2^{f}(2^{-f/2}2^{-k/2})=2^{(f-k)/2} = 2^{-d/2}. \vspace*{-1.5mm}
\]
Thus, the block guessing probability, $2^{-d}$, equals the square of the Bhattacharyya coefficient, $2^{-d}$, between the posterior of $S'$ given $X_{\mathcal{E}}$ and the uniform distribution.
For the standard BEC, the same equivalence can be obtained by averaging this result over the error locations in $\mathcal{E}$.
In Section~\ref{sec:cc_psc}, the analagous result will be considered for the case of PSC-BSC duality.



\section{Brief Review of Quantum Theory}
\label{sec:background}


The central concept of this paper is channel duality, whose formulation for general CQ channels requires somewhat sophisticated techniques from quantum information theory.
However, we focus on the special case of the duality between the classical BSC and the quantum PSC.
Thus, we mainly use linear algebra and some group theory to present and discuss our results.
We do start, however, with a brief review of the quantum theory we will need for our derivations.
It is worth noting that we treat quantum theory from a purely mathematical perspective where each operation is defined without reference to how it might be implemented physically.

\subsection{Dirac Notation and Pure Quantum States}
\label{sec:dirac}

First, we translate the common linear algebraic notation for vectors into the convenient Dirac (or ``bra-ket'') notation used in quantum information theory.

Let $N = 2^n$ for some integer $n \geq 1$ and define $[n] \coloneqq \{1,2,\ldots,n\}$.
We know that a complex length-$N$ vector $\psi \in \mathbb{C}^N$ can be expressed in terms of the standard basis vectors $\{ e_v \in \mathbb{C}^N, v \in \mathbb{Z}_2^n \}$ as $\psi = \sum_{v \in \mathbb{Z}_2^n} \alpha_v e_v$, where $\alpha_v \in \mathbb{C}$, $\mathbb{Z}_2 \coloneqq \{ 0,1 \}$, and $e_v$ contains zeros everywhere except in the entry indexed by $v$.
For $n = 1$, the two basis vectors $e_0$ and $e_1$ are denoted by $\dket{0}$ and $\dket{1}$, respectively, in the Dirac notation.
These are to be read as ``ket $0$'' and ``ket $1$'', respectively, and they are length $N = 2$ column vectors.
Their conjugate transposes are denoted by $\dbra{0} \coloneqq \dket{0}^{\dagger}$ and $\dbra{1} \coloneqq \dket{1}^{\dagger}$, respectively, which are to be read as ``bra $0$'' and ``bra $1$''.
This naming was chosen so that the inner product $\dbraket{i}{j} = \delta_{ij}$, where $i,j \in \{0,1\}$, appears like a bracket (``braket'').
Therefore, any length $N = 2$ complex vector $\psi$ can be expressed as (``ket psi'')
\begin{align}
\psi \equiv \dket{\psi} = \alpha_0 \dket{0} + \alpha_1 \dket{1} = \alpha_0 \begin{bmatrix} 1 \\ 0 \end{bmatrix} + \alpha_1 \begin{bmatrix} 0 \\ 1 \end{bmatrix} = \begin{bmatrix} \alpha_0 \\ \alpha_1 \end{bmatrix} \in \mathbb{C}^2.
\end{align}

The basic unit of quantum information (based on two-level systems) is a \emph{quantum bit} or, simply, a \emph{qubit}.
A qubit that is in a deterministic state is called a (single-qubit) \emph{pure (quantum) state}.
Mathematically, such a qubit ($n = 1$) pure state is simply a unit vector in $\mathbb{C}^2$, which means it can be represented as $\dket{\psi}$ as above with the additional constraint that $|\alpha_0|^2 + |\alpha_1|^2 = 1$.
This normalization constraint is called \emph{Born's rule} and it arises from the measurement postulate of quantum mechanics as we will see shortly.
For $n \geq 1$ qubits, the standard basis vectors $e_v$ are denoted by kets $\dket{v} = e_v = \dket{v_1} \otimes \dket{v_2} \otimes \cdots \otimes \dket{v_n}$, where $v_i \in \mathbb{Z}_2$ for $i \in [n]$ and $\otimes$ denotes the Kronecker product.
Hence, a general $n$-qubit pure state is represented as
\begin{align}
\dket{\psi} = \sum_{v \in \mathbb{Z}_2^n} \alpha_v \dket{v} \in \mathbb{C}^N,\ \ \text{where}\ \ |\dbraket{\psi}|^2 = \sum_{v \in \mathbb{Z}_2^n} |\alpha_v|^2 = 1.
\end{align}
This set of standard basis vectors is called the \emph{comptational basis} of a quantum system.
As might be apparent already, if $\dket{\phi} = \sum_{v \in \mathbb{Z}_2^n} \beta_v \dket{v}$ is another pure state, then its inner product (or overlap) with $\dket{\psi}$ is given by $\dbraket{\phi}{\psi} = \sum_{v \in \mathbb{Z}_2^n} \beta_v^* \alpha_v = \dbraket{\psi}{\phi}^*$.
Sometimes, this inner product is referred to as the \emph{overlap} between the two states.

\subsection{Measurements of Pure States}
\label{eq:measurements}

The only way to obtain classical information about a quantum system is \emph{measurement}.
The simplest kind of measurement is called a \emph{von Neumann} (or projective) measurement.
A von Neumann measurement is defined by a set of orthogonal projectors $\{ \Pi_i,\ i \in [M] \}$ such that $\Pi_i \Pi_j = \delta_{ij} \Pi_i$ and $\sum_{i = 1}^M \Pi_i = I_N$, where $I_N$ is the $N \times N$ identity matrix.
If the state of the underlying quantum system is $\dket{\psi}$, then this measurement produces a classical label $i \in [M]$ with a certain probability and also causes the projector $\Pi_i$ to be applied to the state, as stated by the measurement postulate of quantum mechanics.
By the \emph{Born rule}, this event happens with the probability $p_i = \dbra{\psi} \Pi_i \dket{\psi}$~\cite{Wilde-2013}.
The postulate also states that, from the classical label $i$, we can be sure that the new state of the system is given by $\dket{\psi_i} = \Pi_i \dket{\psi}/\sqrt{p_i}$.
Hence, the measurement produces a random classical label $i$ that indicates the subspace (i.e., range of $\Pi_i$) upon which the initial state has been projected.
From a linear algebraic standpoint, a projective measurement involves splitting the vector space $\mathbb{C}^N$ into several orthogonal subspaces, and then performing a random projection of the system's state onto one of these subspaces according to the Born rule.

Note that this change to the original state is a distinguishing feature of quantum systems.
For classical systems such as a received waveform in standard wireless communications, performing a ``measurement'' such as an inner product with a locally generated waveform, does not alter the received waveform.
In this paper, though we are not concerned about quantum error correction, we will use such measurements to identify one of several classical messages transmitted through a CQ channel.

From the definition of the measurement projectors, it is evident that $\sum_{i = 1}^M p_i = \sum_{i = 1}^M \dbra{\psi} \Pi_i \dket{\psi} = \dbra{\psi} \left( \sum_{i = 1}^M \Pi_i \right) \dket{\psi} = 1$ and $\dbraket{\psi_i} = \dbra{\psi} \Pi_i \dket{\psi}/p_i = 1$, as necessary.
A special case of the von Neumann measurement is the scenario where all the projectors have rank $1$, i.e., $\Pi_i = \dketbra{\phi_i}$ for some set of orthogonal quantum states $\{ \dket{\phi_i},\ i \in [M] \}$ and $M=N$.
More generally, an arbitrary quantum measurement is described by a \emph{positive operator-valued measure (POVM)}~\cite{Wilde-2013}.
However, we will only require projective measurements in this paper.

\subsection{Mixed Quantum States}
\label{sec:mixed_states}

A quantum system can also be in a random state.
In general, it can be in one of several states $\{ \dket{\psi_m},\ m \in [T] \}$ with associated probabilities $p_m$.
For such a ``bag of states'' model, a succinct description of the state of the system is given by its \emph{density matrix} 
\begin{align}
\rho \coloneqq \sum_{m = 1}^{T} p_m \dketbra{\psi_m} \in \mathbb{C}^{N \times N}.
\end{align}
It is easy to verify that this matrix is positive semi-definite and has trace $1$; in fact, any operator that satisfies these properties is a valid density matrix for some quantum system.
Note that the above expansion is not an eigendecomposition unless the state vectors are orthogonal. 
Clearly, its eigenvectors and eigenvalues provide another quantum system described by the same density matrix.
Therefore, the density matrix of a given quantum system is uniquely defined but the interpretation of a density matrix as a mixture of pure states is not necessarily unique.
The density matrix corresponds to a pure state if and only if $T = 1$; otherwise it corresponds to a \emph{mixed (quantum) state}.
Note that a general density matrix is commonly described as a mixed state in the literature.

Consider the effect of a von Neumann measurement $\{ \Pi_i \}$ on a system described by the mixed state $\rho = \sum_{m = 1}^{T} p_m \dketbra{\psi_m}$.
Let $p_{i|m}$ denote the conditional probability of measurement outcome $i$ given that the system is in pure-state $\dket{\psi_m}$ and $p_{m|i}$ denote the posterior probability that the system is in pure-state $\dket{\psi_m}$ given measurement outcome $i$.
If the outcome is $i$, then the overall post-measurement (mixed) state is
\begin{align}
\label{eq:rho_i_msmt}
\rho_i & = \sum_{m = 1}^{T} p_{m|i} \dketbra{\psi_m^i} = \frac{\Pi_i\, \rho\, \Pi_i}{p_i},\\ 
\text{where}\ \ p_{m|i} & = \frac{p_{i|m} p_m}{p_i}, \ p_{i|m} = \dbra{\psi_m} \Pi_i \dket{\psi_m},\ \dket{\psi_m^i} = \frac{\Pi_i \dket{\psi_m}}{\sqrt{p_{i|m}}}.
\end{align}
Computing the trace on both sides of~\eqref{eq:rho_i_msmt} and using the cyclic property of the trace shows that
\begin{align}
\label{eq:density_msmt_prob}
p_i & = \sum_{m' = 1}^{T} p_{i|m'} p_{m'} = \text{Tr}\left[ \Pi_i \rho \right].
\end{align}
Therefore, the density matrix encodes all the necessary information about the system in order to track its evolution through arbitrary quantum processes, i.e., unitary operations and measurements.
It is important to keep in mind that the density matrix description only traces the system's evolution \emph{on average}.
So, if one cares about the evolution of certain specific constituents, i.e., certain pure states in a given decomposition of the density matrix, then one needs to evolve those pure states separately.

\subsection{Linear Codes and their Complements}
\label{sec:linear_codes}

In this section, we review a particular perspective on linear codes and their complements as described by Renes in~\cite{Renes-it18}, but by replacing parity-check matrices with generator matrices.
A binary linear code $\MCC \colon [n,k]$ and its dual code $\MCC^{\perp} \colon [n,n-k]$ can be related through their complementary codes $\MCC^\top \colon [n,n-k]$ and $\MCC^\topbot \colon [n,k]$, respectively.
To see this, let us define a code and its dual in a different way. 
First, define a invertible linear transformation $A$ from $\mathbb{Z}_{2}^{n}$ to itself\footnote{This is analogous to the matrix $\Ag$ in Section~\ref{sec:bec_duality} but we rename its constituents for more clarity.}. 
Then we can regard the first $k$ rows of $A$ as a generator matrix $G_{\MCC} \in \mathbb{Z}_{2}^{k \times n}$ of $\mathcal{C}$. 
The remaining (last) $(n-k)$ rows of $A$ form a generator matrix $G_{\MCC^{\top}} \in \mathbb{Z}_{2}^{(n-k) \times n}$ of $\MCC^\top$. 
The code $\MCC^\top$ (resp. $\MCC$) is called the complement of $\MCC$ (resp. $\MCC^\top$) because $\MCC^\top \oplus \MCC = \mathbb{Z}_2^n$. 

Next, define $B \coloneqq \left(A^{-1}\right)^{T}$ and define $G_{\MCC^\topbot}$ and $G_{\MCC^\perp}$ to be its first $k$ rows and last $(n-k)$ rows, respectively.
Then, $G_{\MCC^\perp}$ is a generator matrix of the dual code $\MCC^{\perp}$ and $G_{\MCC^\topbot}$ is a generator matrix of the dual-complement $\MCC^{\topbot}$. 
Hence, we have the following picture in terms of the generator matrices of these codes:
\begin{align}
A = 
\begin{bmatrix}
G_{\MCC}\\
G_{\MCC^\top}
\end{bmatrix} 
\quad ; \quad 
B= \left(A^{-1}\right)^{T} = 
\begin{bmatrix}
G_{\MCC^\topbot}\\
G_{\MCC^\perp}
\end{bmatrix}
\Rightarrow A^{-1} = \left[ G_{\MCC^\topbot}^{T},\ G_{\MCC^\perp}^{T} \right].
\end{align}
Since $A B^T = A A^{-1} = I_n$, denoting by $O$ the matrix with all zeros, we have
\begin{align}
\label{eq:H_relations}
I_n = A B^T = 
\begin{bmatrix}
G_{\MCC} G_{\MCC^\topbot}^{T} & G_{\MCC} G_{\MCC^\perp}^{T}\\
G_{\MCC^\top} G_{\MCC^\topbot}^{T} & G_{\MCC^\top} G_{\MCC^\perp}^{T}
\end{bmatrix} \Rightarrow 
G_{\MCC} G_{\MCC^\perp}^{T} & = O_{k \times (n-k)}\ (\MCC \perp \MCC^{\perp}), \ G_{\MCC^\top} G_{\MCC^\topbot}^{T} = O_{(n-k) \times k}\ (\MCC^{\top} \perp \MCC^{{\scriptstyle \topbot}}), \nonumber \\
  G_{\MCC} G_{\MCC^\topbot}^{T} & = I_{k}\ \text{and}\ G_{\MCC^\top} G_{\MCC^\perp}^{T} = I_{n-k}.
\end{align}
We will use this organization of codes frequently in the rest of the paper.


\subsection{Pure-State Channel (PSC)}
\label{sec:psc}

The pure-state CQ channel can be described by the mapping $W^{\psc(\theta)}\colon x \in \mathbb{Z}_2 \mapsto \dket{(-1)^x\theta} \equiv \dketbra{(-1)^x\theta}$, where 
\begin{align}
\dket{(-1)^x\theta} \coloneqq \cos\frac{\theta}{2} \dket{0} + (-1)^x \sin\frac{\theta}{2} \dket{1},\ \ \theta \in \left[ 0, \frac{\pi}{2} \right].
\end{align}
The overlap between the states is $\dbraket{-\theta}{\theta} = \cos^2 \frac{\theta}{2} - \sin^2 \frac{\theta}{2} = \cos\theta$.
Hence, when $\theta = \pi/2$ the PSC outputs one of two orthogonal states which can be detected perfectly by performing the rank-$1$ von Neumann measurement $\{ \dketbra{+}, \dketbra{-} \}$, where $\dket{\pm} \coloneqq \frac{\dket{0} \pm \dket{1}}{\sqrt{2}}$.
For $\theta \neq \pi/2$, any measurement will incur a non-zero probability of error in detecting which of the two states was output by the channel.
The optimal probability of error is achieved by the Helstrom measurement~\cite{Helstrom-jsp69,Helstrom-ieee70}, which in this case turns out to be the aforesaid measurement.
So, for any $\theta$, the optimal measurement $\{ \dketbra{+}, \dketbra{-} \}$ achieves the probability of error $P_{\text{Hel}}^{\psc} = \frac{1}{2} \left[ 1 - \sqrt{1 - |\dbraket{-\theta}{\theta}|^2} \right] = (1 - \sin\theta)/2$.
Hence, the PSC combined with this Helstrom measurement induces the binary symmetric channel \bsc($P_{\text{Hel}}^{\psc}$).

However, the Holevo capacity of the PSC, i.e., its capacity for transmitting classical information reliably, is significantly higher than that of this induced BSC.
Therefore, if we use a classical length-$n$ code to communicate over the PSC, then measuring each of the $n$ output qubits and post-processing them classically is suboptimal.
This suboptimality is true in the block error probability sense as well.
In this paper, we will derive the optimal block error probability for channel coding over the PSC using the \emph{square root measurement} (SRM) that we discuss shortly.

Observe that the two possible output states of the PSC satisfy a symmetry: $\dket{-\theta} = Z \dket{\theta}$, where $Z \coloneqq \begin{bsmallmatrix} 1 & 0 \\ 0 & -1 \end{bsmallmatrix}$ is the Pauli $Z$ operator.
For a binary vector $b = [b_1,b_2,\ldots,b_n]$ we define $Z(b) \coloneqq Z(b_1) \otimes Z(b_2) \otimes \cdots \otimes Z(b_n)$, where $Z(b_i) \coloneqq Z^{b_i}$.
Then, when $b \in \mathbb{Z}_2^n$ is transmitted over $n$ uses of the PSC, the output state is given by $Z(b) \dket{\theta}^{\otimes n}$.

On the single qubit computational basis states, $Z$ acts as $Z \dket{0} = \dket{0}, Z \dket{1} = - \dket{1}$.
So, on an $n$-qubit computational basis state $\dket{v}, v \in \mathbb{Z}_2^n$, the action of $Z(b)$ is given by $Z(b) \dket{v} = (-1)^{vb^T} \dket{v}$.
Hence, we can express $Z(b) = \sum_{v \in \mathbb{Z}_2^n} (-1)^{vb^T} \dketbra{v}$.

\subsection{Von Neumann Entropy}

For information-theoretic quantities involving quantum systems, we adopt the notation used in~\cite{Wilde-2013} because it highlights similarity with related classical quantities.
For a quantum system $Y$ described by an $N \times N$ density matrix $\rho^Y$, the Von Neumann entropy of $Y$ is defined to be
$$ H(Y)_{\rho^Y} \triangleq - \sum_{i=1}^N \lambda_i \log \lambda_i , $$
where $\lambda_1,\ldots,\lambda_N$ are the eigenvalues of $\rho^Y$ and $\lambda \log \lambda \triangleq 0$ for $\lambda = 0$.
Of course, this is simply the Shannon entropy of the eigenvalues of $\rho^Y$.

For a pair of quantum systems $X$ and $Y$, the joint density matrix $\rho^{XY}$ is indexed by $(x,y)$ pairs so that $\rho^{XY}_{(x,y),(x',y')}$ is an entry of this matrix.
These $(x,y)$ pairs implicitly represent kronecker products of standard basis vectors for the two quantum systems.
Thus, the density matrices of the individual systems, which essentially describe the ``marginals" of the joint system, are defined using the partial trace operations
$$ \rho^X_{x,x'} = \text{Tr}_Y \left[\rho^{XY}\right] \triangleq \sum_y \rho^{XY}_{(x,y),(x',y)}, \quad\quad \rho^Y_{y,y'} = \text{Tr}_X \left[\rho^{XY}\right] \triangleq \sum_x \rho^{XY}_{(x,y),(x,y')}. $$
With this, the Von Neumann conditional entropy and mutual information are defined by evaluating the classical formulas with the Von Neumann entropy to get
\begin{align*}
H(X|Y)_{\rho^{XY}} &\triangleq H(X,Y)_{\rho^{XY}} - H(Y)_{\rho^{Y}} \\
I(X;Y)_{\rho^{XY}} &\triangleq H(X)_{\rho^{X}} + H(Y)_{\rho^{Y}} - H(X,Y)_{\rho^{XY}} = H(X)_{\rho^{X}} - H(X|Y)_{\rho^{XY}}.
\end{align*}
These quantities also reduce to their classical counterparts when the associated density matrices are diagonal.
For a comprehensive discussion of these quantities and their operational interpretations, see~\cite{Wilde-2013}.

\section{Duality Between Channel Coding and Secret Communications}
\label{sec:duality_coding_secrecy}

\subsection{Block Error Rate of Channel Coding on the PSC}
\label{sec:cc_psc}

For a $M$-ary hypothesis testing problem with candidate states $\{ \rho_i,\ i \in [M] \}$, the minimum Bayes cost is given by
\begin{align}
\label{eq:bayes_cost}
C^* = \sum_{j=1}^M \tr{\hat{\Pi}_j \rho_j'}, \quad \rho_j' \coloneqq \sum_{i=1}^M p_i C_{ji} \rho_i,
\end{align}
where $C_{ji}$ is the cost associated to deciding $\rho_j$ when the truth is $\rho_i$, and $\{ \hat{\Pi}_j,\ j \in [M] \}$ is the optimal POVM.
For the transmission of an $[n,k]$ binary linear code $\MCC$ on $\psc(\theta)$, the minimum block error probability $P_e$ equals $C^*$ using the hypothesis testing problem with $C_{ji} = 1-\delta_{ji}$ and $\rho_i = \dketbra{\phi_i}$, where $\dket{\phi_i}$ is the result of transmitting the $i$-th codeword of $\MCC$ over\footnote{As in Section~\ref{sec:bec_duality}, one could add a coset vector $sF$ to the codeword and then assume that the receiver is informed of this vector. But, we avoid this here because it only complicates the problem and it is easily removed at the PSC output by applying $Z(sF)$ deterministically.} the $\psc(\theta)$.
This problem satisfies the \emph{geometrically uniform (GU)} state set criterion of Eldar and Forney~\cite{Eldar-it00}.
The criterion is that there is a generator state $\dket{\phi}$ and an abelian group $\mathcal{G}$ such that each $\dket{\phi_i}$ can be written as $\dket{\phi_i} = U_i \dket{\phi}$ for some $U_i \in \mathcal{G}$.
For this case, by the PSC symmetry mentioned in Section~\ref{sec:psc}, we have $\dket{\phi} = \dket{\theta}^{\otimes n}$ and $\mathcal{G} = \{ Z(c),\ c \in \MCC \}$.
Assuming each codeword is equally likely, the results in~\cite{Eldar-it00} show that the  \emph{square-root measurement (SRM)} (also called the \emph{pretty-good measurement (PGM)}) is the optimal POVM $\{ \hat{\Pi}_j,\ j \in [M] \}$.

\begin{definition}
\label{def:srm}
The elements of the SRM POVM are $\hat{\Pi}_j = \dketbra{\psi_j}$, where $\dket{\psi_j}$ is the $j$-th column of the SRM matrix
\begin{align}
\Psi \coloneqq \Phi \left( \left( \Phi^{\dagger} \Phi \right)^{1/2} \right)^{-1} \in \mathbb{C}^{2^n \times 2^k},
\end{align}
the columns of $\Phi$ are $\{ \dket{\phi_{i-1}} ,\ i \in [2^k] \}$, and the inverse is actually the Moore-Penrose pseudo-inverse.
In terms of the compact singular value decomposition (SVD), $\Phi = U \Sigma V^\dagger$, one can also write $\Psi = U V^\dagger$.
\end{definition}

However, for the transmission of a binary linear code over the PSC, the Gram matrix $\Phi^{\dagger} \Phi$ has full rank because $\Phi$ has full column rank.
Note that the columns of $\Phi$ can be written as $\dket{\phi_i} = Z(c_i) \dket{\theta}^{\otimes n}$ for some indexing of the codewords $c_i \in \MCC$.
Hence, if columns of $\Phi$ were linearly dependent, then that would mean that $\dket{\theta}^{\otimes n}$ (which has no zero entries for non-extremal $\theta$) is in the null space of a sum of $Z(c_i)$'s (which is a diagonal matrix).
This is clearly impossible.

\begin{lemma}
For channel coding over $\psc(\theta)$ with an $[n,k]$ binary linear code, if $P_e$ is the block error probability, then
\begin{align}
P_e & = \frac{1}{2^k} \sum_{j \in \{0,1\}^k} \sum_{\substack{i \in \{0,1\}^k\\i \neq j}} \left| \dbraket{\psi_j}{\phi_i} \right|^2 = \frac{1}{2^k} \sum_{j \in \{0,1\}^k} \sum_{\substack{i \in \{0,1\}^k\\i \neq j}} \left| (\Psi^{\dagger} \Phi)_{ji} \right|^2.
\end{align}
\end{lemma}
\begin{IEEEproof}
In this setting, we choose the cost function to be $C_{ji} \coloneqq 1 - \delta_{ji}$ where $\delta_{ji}$ denotes the Kronecker delta function. 
Thus, the optimal cost $\hat{C}$ equals the error probability $P_e$ and this implies
\begin{align}
\rho_j' = \sum_{\substack{i=0\\i \neq j}}^{2^k-1} p_i \rho_i = \frac{1}{2^k} \sum_{\substack{i=0\\i \neq j}}^{2^k-1} \dketbra{\phi_i} = \frac{1}{2^k} \sum_{\substack{i=0\\i \neq j}}^{2^k-1} Z(c_i) \dketbra{\theta}^{\otimes n} Z(c_i).
\end{align}
Similarly, the optimal POVM elements are $\hat{\Pi}_j = \dketbra{\psi_j} = \Psi \dketbra{j} \Psi^{\dagger}$ for $j \in \{0,1\}^k$.
Therefore,
\begin{align}
\tr{\hat{\Pi}_j \rho_j'} & = \frac{1}{2^k} \sum_{\substack{i \in \{0,1\}^k\\i \neq j}} \tr{ \Psi \dketbra{j} \Psi^{\dagger} Z(c_i) \dketbra{\theta}^{\otimes n} Z(c_i) } \\
  & = \frac{1}{2^k} \sum_{\substack{i \in \{0,1\}^k\\i \neq j}} \tr{ \dbra{j} \Psi^{\dagger} Z(c_i) \dketbra{\theta}^{\otimes n} Z(c_i) \Psi \dket{j} } \\
  & = \frac{1}{2^k} \sum_{\substack{i \in \{0,1\}^k\\i \neq j}} \left| \dbra{j} \Psi^{\dagger} Z(c_i) \dket{\theta}^{\otimes n} \right|^2 \\
  & = \frac{1}{2^k} \sum_{\substack{i \in \{0,1\}^k\\i \neq j}} \left| (\Psi^{\dagger} \Phi)_{ji} \right|^2 \\
  & = \frac{1}{2^k} \sum_{\substack{i \in \{0,1\}^k\\i \neq j}} \left| \dbraket{\psi_j}{\phi_i} \right|^2.
%
\end{align}
Substituting this expression in~\eqref{eq:bayes_cost} completes the proof.
Though the final expression can be written immediately by looking at the form of $\hat{\Pi}_j$ and $\rho_j'$, the intermediate steps reveal more information about the form of the inner products.
\end{IEEEproof}

The remainder of this section is devoted to calculating $P_e$ in closed form, using tools developed in~\cite{Eldar-it00}.
First, our candidate states are GU with generator $\dket{\theta}^{\otimes n}$ and the abelian group $\mathcal{G} = \{ Z(c), \ c \in \MCC \}$.
If we express $c = m G$ with respect to some generator matrix $G$ for $\MCC$, where $m \in \{0,1\}^k$ are arbitrary message vectors, then we see that $Z(m_1 G) Z(m_2 G) = Z(c_1) Z(c_2) = Z(c_1 \oplus c_2) = Z((m_1 \oplus m_2) G)$, which clearly means $\mathcal{G} \cong \mathbb{Z}_2^k = \{0,1\}^k$.
The isomorphism is given explicitly by $Z(mG) \leftrightarrow m$. 
Note that, given an invertible $A \in \mathbb{Z}_2^{n \times n}$, we can set 
$G = G_{\MCC}$ as per the discussion in Section~\ref{sec:linear_codes}.

\begin{definition}
\label{def:fourier}
The Fourier transform of a function $\varphi \colon \mathbb{Z}_2^k \rightarrow \mathbb{C}$ is the function $\hat{\varphi} \colon \mathbb{Z}_2^k \rightarrow \mathbb{C}$ defined by
\begin{align}
\hat{\varphi}(h) = \frac{1}{\sqrt{2^k}} \sum_{g \in \mathbb{Z}_2^k} (-1)^{hg^T} \varphi(g),
\end{align}
where $hg^T = \sum_{i=1}^k h_i g_i$ (mod $2$) is the binary inner product between $h$ and $g$.
The Fourier transform matrix $\mathcal{F}$ is given by $\mathcal{F}_{gh} = \frac{1}{\sqrt{2^k}} (-1)^{gh^T}$, where the rows and columns are indexed by $g,h \in \mathbb{Z}_2^k$.
\end{definition}

\begin{definition}
\label{def:overlap_fn}
Given a linear code $\MCC$ that is transmitted over $\psc(\theta)$, define the overlap function
\begin{align}
s(g) \coloneqq \dbra{\theta}^{\otimes n} Z(c_g) \dket{\theta}^{\otimes n} = \left( \cos\theta \right)^{w_H(c_g)},
\end{align}
where $w_H(c_g)$ is the Hamming weight of the codeword $c_g \coloneqq g G_{\MCC} \in \MCC, g \in \mathbb{Z}_2^k$.
By Definition~\ref{def:fourier}, its Fourier transform is
\begin{align}
\label{eq:shat_h}
\hat{s}(h) = \frac{1}{\sqrt{2^k}} \sum_{g \in \mathbb{Z}_2^k} (-1)^{hg^T} \left( \cos\theta \right)^{w_H(c_g)}.
\end{align}
\end{definition}

\begin{remark} \label{rem:diff_theta}
The results in this section also extend to the case where the $i$-th bit is transmitted over a PSC with parameter $\theta_i$.
In that case, the term $\left( \cos\theta \right)^{w_H (c_g)}$ changes to
$ \prod_{i: [c_g]_i = 1} \cos \theta_i $ and the derivations can be adjusted accordingly.
However, we restrict the full exposition to the simpler case where $\theta_i = \theta$ for all $i$.
\end{remark}

We will show later that $\{ 2^{-k/2} \hat{s}(h),\ h \in \mathbb{Z}_2^k \}$ forms the posterior distribution for secret communications over the binary symmetric channel $\bsc\left( \frac{1-\cos\theta}{2} \right)$, when cosets of $\MCCd$ are used to send secure messages.

\begin{lemma}
The function $\hat{s}(h)$ satisfies $\sum_{h \in \mathbb{Z}_2^k} 2^{-k/2} \hat{s}(h) = 1$.
\end{lemma}
\begin{IEEEproof}
We calculate
\begin{align}
\frac{1}{\sqrt{2^k}} \sum_{h \in \mathbb{Z}_2^k} \hat{s}(h) & = \frac{1}{\sqrt{2^k}} \sum_{h \in \mathbb{Z}_2^k} \frac{1}{\sqrt{2^k}} \sum_{g \in \mathbb{Z}_2^k} (-1)^{hg^T} \left( \cos\theta \right)^{w_H(c_g)} \\
  & = \frac{1}{2^k} \sum_{g \in \mathbb{Z}_2^k} \left( \cos\theta \right)^{w_H(c_g)} \cdot \left( \sum_{h \in \mathbb{Z}_2^k} (-1)^{hg^T} \right) \\
  & \overset{\text{(a)}}{=} \frac{1}{2^k} \left( \cos\theta \right)^{w_H(0)} \cdot \left( 2^k \, \mathbb{I}(g=0) \right) \\
  & = 1. 
\end{align}
In step (a), we used the fact that the inner summation vanishes unless $g = 0$.
\end{IEEEproof}

Now, using the above definitions, we will state a key result that enables us to calculate $P_e$ in closed-form.

\begin{theorem}[Adapted from Eldar and Forney~\cite{Eldar-it00}]
Consider the transmission of an $[n,k]$ binary linear code $\MCC$ over the channel $\psc(\theta)$.
The codeword matrix $\Phi$ and the SRM matrix $\Psi$ satisfy $\Psi^{\dagger} \Phi = \mathcal{F} \overline{\Sigma} \mathcal{F}^{\dagger}$, where $\overline{\Sigma}$ is a $2^k \times 2^k$ diagonal matrix with diagonal elements 
\begin{align}
\sigma(h) \coloneqq 2^{k/4} \sqrt{\hat{s}(h)}, \ \ h \in \mathbb{Z}_2^k.
\end{align}
\end{theorem}
%
%

Since $\Psi^{\dagger} \Phi = \mathcal{F} \overline{\Sigma} \mathcal{F}^{\dagger}$ is Hermitian, we also observe that $(\Psi^{\dagger} \Phi)_{ji} = (\Psi^{\dagger} \Phi)_{ij}^* \Rightarrow \left| (\Psi^{\dagger} \Phi)_{ji} \right|^2 = \left| (\Psi^{\dagger} \Phi)_{ij} \right|^2$, and hence
\begin{align}
\label{eq:Pe_reduced}
P_e & = \frac{1}{2^k} \sum_{j \in \mathbb{Z}_2^k} \sum_{\substack{i \in \mathbb{Z}_2^k\\i \neq j}} \left| \dbraket{\psi_j}{\phi_i} \right|^2 = \frac{1}{2^k} \sum_{j \in \mathbb{Z}_2^k} \sum_{\substack{i \in \mathbb{Z}_2^k\\i > j}} 2 \left| \dbraket{\psi_j}{\phi_i} \right|^2,
\end{align}
where $i > j$ should be interpreted according to the decimal equivalent of $i$ and $j$.
Using the expressions in~\cite{Eldar-it00}, the columns of $\Psi$ are given by $\{ \dket{\psi_g},\, g \in \mathbb{Z}_2^k \}$, where
\begin{align}
\dket{\psi_g} & = \frac{1}{\sqrt{2^k}} \sum_{h \in \mathbb{Z}_2^k} (-1)^{gh^T} \frac{1}{\sigma(h)} \mathbb{I}(\sigma(h) \neq 0) \frac{1}{\sqrt{2^k}} \sum_{f \in \mathbb{Z}_2^k} (-1)^{hf^T} Z(c_f) \dket{\theta}^{\otimes n} \\
\label{eq:psi_g}
  & = \frac{1}{2^k} \sum_{h,f \in \mathbb{Z}_2^k} \mathbb{I}(\sigma(h) \neq 0) \frac{(-1)^{h (f \oplus g)^T}}{\sigma(h)} Z(c_f) \dket{\theta}^{\otimes n}.
\end{align}
Hence, to compute the inner products $\left| \dbraket{\psi_j}{\phi_i} \right|$ in $P_e$, we need to calculate $\sigma(h)$ or, equivalently, $\hat{s}(h)$ for all $h \in \mathbb{Z}_2^k$. \\

\noindent \textbf{Factor Graph Duality Enables Calculation of Block Error Probability} \\

We will now introduce the indicator function of $\MCC$ in $\hat{s}(h)$, in order to apply a factor graph duality~\cite{Hartmann-it76,Forney-it01,Forney-arxiv11} that produces the indicator function of $\MCC^{\perp}$ and simplifies the calculation of $\hat{s}(h)$. 
For this, let us embed $s(g)$ in $\mathbb{Z}_2^n$ by the definition 
\begin{align}
s'(x) \coloneqq \mathbb{I}(x \in \MCC) \left( \cos\theta \right)^{w_H(x)},\ x \in \mathbb{Z}_2^n.
\end{align}
Then, the Fourier transform over $\mathbb{Z}_2^n$ produces
\begin{align}
\label{eq:embed_s}
\hat{s}'(y) = \frac{1}{\sqrt{2^n}} \sum_{x \in \mathbb{Z}_2^n} (-1)^{yx^T} \mathbb{I}(x \in \MCC) \left( \cos\theta \right)^{w_H(x)}.
\end{align}

\begin{remark}
\label{rem:shatp_y}
We immediately observe that, if we express $y = y_{\MCC^{\topbot}} + x_{\MCCd}$ for some unique $y_{\MCC^{\topbot}} \in \MCC^{\topbot}$ and $x_{\MCCd} \in \MCCd$, then $\hat{s}'(y) = \hat{s}'(y_{\MCC^{\topbot}})$ since $yx^T = y_{\MCC^{\topbot}} x^T$ for all $x \in \MCC$.
\end{remark}

Now, we see that the sum effectively happens over $\mathbb{Z}_2^k$ due to the presence of the indicator function, but the factor $(-1)^{yx^T}$ does not exactly map to $(-1)^{hg^T}$ since the latter is only taken over the ``message'' part of the codewords $x,y$ assuming a systematic encoding.
Hence, we need to make these coefficients $\hat{s}'(y)$ usable exactly in place of $\hat{s}(h)$.
This means we need to ensure that whenever $x,y \in \MCC$, the exponent satisfies $xy^T \equiv \sum_{i=1}^k x_i y_i\ (\text{mod}\ 2)$, where the first $k$ bits of $x$ and $y$ can be identified with the ``message'' vectors $g$ and $h$ above.
Since $x = x_{1:k} G_{\MCC}$ and $y = y_{1:k} G_{\MCC}$, this means we need $x_{1:k} y_{1:k}^T \equiv x_{1:k} G_{\MCC} G_{\MCC}^T y_{1:k}^T\ (\bmod\ 2)$, which implies we need $G_{\MCC} G_{\MCC}^T \equiv I_k\ (\text{mod}\ 2)$.
This is clearly not possible for all codes $\MCC$ and a simple counterexample is an even code $\MCC$.
In order to circumvent this problem, we exploit the alternative perspective of linear codes in Section~\ref{sec:linear_codes}.


\begin{lemma} \label{lem:shat_shatp}
Let $y_h=h G_{\MCC^\topbot}$ denote the codeword in $\MCC^{\topbot}$ corresponding to the message $h \in \mathbb{Z}_2^k$, i.e., $y_h = [h , 0^{n-k}] B$.
Then, the functions $\hat{s}(h)$ and $\hat{s}'(y_h)$ are related by
\begin{align}
\hat{s}(h) = \sqrt{2^{n-k}} \hat{s}'(y_h).
\end{align}
This further implies that $\sigma(h) = 2^{k/4} \sqrt{\hat{s}(h)} = 2^{n/4} \sqrt{\hat{s}'(y_h)}$.
\end{lemma}
\begin{IEEEproof}
Given $y \in \mathbb{Z}_2^n$, express it uniquely as $y = y_h \oplus x_{\MCCd}$ for some $h \in \mathbb{Z}_2^k$ and $x_{\MCCd} \in \MCCd$.
Then, using Remark~\ref{rem:shatp_y},
\begin{align}
\hat{s}'(y) = \hat{s}'(y_h) & = \frac{1}{\sqrt{2^n}} \sum_{x \in \mathbb{Z}_2^n} (-1)^{y_h x^T} \mathbb{I}(x \in \MCC) \left( \cos\theta \right)^{w_H(x)} \\
%
  & = \frac{1}{\sqrt{2^n}} \sum_{x \in \MCC} (-1)^{ y_h B^{-1} B x^T} \left( \cos\theta \right)^{w_H(x)} \\
%
  & = \frac{1}{\sqrt{2^n}} \sum_{x \in \MCC} (-1)^{(y_h B^{-1}) (x A^{-1})^T} \left( \cos\theta \right)^{w_H(x)} \\
%
  & = \frac{1}{\sqrt{2^n}} \sum_{x \in \MCC} (-1)^{\begin{bmatrix} h & 0^{n-k} \end{bmatrix} \begin{bmatrix} g_x^T \\ \left( 0^{n-k} \right)^T \end{bmatrix}} \left( \cos\theta \right)^{w_H(x)} \\
  & = \frac{1}{\sqrt{2^n}} \sum_{x \in \MCC} (-1)^{h g_x^T} \left( \cos\theta \right)^{w_H(x)} \\
  & = \frac{1}{\sqrt{2^n}} \sum_{g \in \mathbb{Z}_2^k} (-1)^{hg^T} \left( \cos\theta \right)^{w_H(g G_{\MCC})}.
\end{align}
Here, $x = g_x G_{\MCC}$, and by comparing with $\hat{s}(h)$ in~\eqref{eq:shat_h} we conclude that $\hat{s}(h) = \sqrt{2^{n-k}} \hat{s}'(y_h)$.
This relation can be induced by
\begin{IEEEeqnarray*}{rCl+x*}
\hat{s}(h) & \coloneqq & \frac{1}{\sqrt{2^{n-k}}} \sum_{x_{\MCCd} \in \MCCd} \hat{s}'(y_h \oplus x_{\MCCd}) = \sqrt{2^{n-k}} \hat{s}'(y_h). & \IEEEQEDhere
\end{IEEEeqnarray*}
\end{IEEEproof}

We will derive a closed-form expression for $\hat{s}(h)$ using the above relation and the following well-known result.

\begin{lemma}[Factor graph duality~\cite{Forney-arxiv11}]
\label{lem:fg_duality}
For binary vectors $x \in \mathbb{Z}_2^n$, given functions $\mu_j \colon \mathbb{Z}_2 \rightarrow \mathbb{R}$ for each index $j \in \{1,2,\ldots,n\}$, and an $[n,k]$ binary linear code $\MCC$, we have
\begin{align}
\sum_{x \in \mathbb{Z}_2^n} \mathbb{I}(x \in \MCC) \prod_{j=1}^n \mu_j(x_j) = \sum_{\hat{x} \in \mathbb{Z}_2^n} 2^{k-n/2} \mathbb{I}(\hat{x} \in \MCC^{\perp}) \prod_{j=1}^n \hat{\mu}_j(\hat{x}_j),
\end{align}
where $\hat{\mu}_j(\hat{z}) \coloneqq \frac{1}{\sqrt{2}} \sum_{z \in \mathbb{Z}_2} (-1)^{\hat{z} z} \mu_j(z)$.
\end{lemma}
\begin{IEEEproof}
See Appendix~\ref{sec:fg_duality}, taken from~\cite{Pfister-fg_duality}, for an algebraic proof rather than the graphical approach in~\cite{Forney-arxiv11}.
\end{IEEEproof}

\begin{lemma} \label{lem:FT_overlap}
Given an $[n,k]$ binary linear code $\MCC$ and the channel $\psc(\theta)$, the Fourier transform $\hat{s}(h), h \in \mathbb{Z}_2^k$, of the overlap function $s(g)$ is given by
\begin{align}
\label{eq:shat_final}
\frac{1}{2^{k/2}} \hat{s}(h)  & = \!\! \sum_{z \in y_h\, \oplus\, \MCC^{\perp}} p^{w_H(z)} (1 - p)^{n - w_H(z)}\ ; \ \ \sum_{h \in \mathbb{Z}_2^k} \frac{\hat{s}(h)}{2^{k/2}} = 1.
\end{align}
Here, $y_h$ denotes the codeword in $\MCC^{\topbot}$ corresponding to the message $h \in \mathbb{Z}_2^k$, i.e., 
$y_h = [h, 0^{n-k}] B$ (see Section~\ref{sec:linear_codes}).
\end{lemma}
\begin{IEEEproof}
Using Lemma~\eqref{lem:shat_shatp} and setting $y = y_h$, we can now write
\begin{IEEEeqnarray}{rCl+x*}
\hat{s}'(y_h) & = & \frac{1}{\sqrt{2^n}} \sum_{x \in \mathbb{Z}_2^n} (-1)^{h g_x^T} \mathbb{I}(x \in \MCC) \left( \cos\theta \right)^{w_H(x)} & \\
  & = & \frac{1}{\sqrt{2^n}} \sum_{x \in \mathbb{Z}_2^n} (-1)^{y x^T} \mathbb{I}(x \in \MCC) \left( \cos\theta \right)^{w_H(x)} & \\
  & = & \sum_{x \in \mathbb{Z}_2^n} \mathbb{I}(x \in \MCC) \prod_{j=1}^n \frac{1}{\sqrt{2}} (-1)^{y_j x_j} \left( \cos\theta \right)^{x_j} & \\
  & = & \sum_{x \in \mathbb{Z}_2^n} \mathbb{I}(x \in \MCC) \prod_{j=1}^n \mu_j(x_j),
\end{IEEEeqnarray}
where $\mu_j(z) = \frac{1}{\sqrt{2}} \left( (-1)^{y_j} \cos\theta \right)^{z}$.
To apply Lemma~\ref{lem:fg_duality}, we compute
\begin{IEEEeqnarray}{rCl+x*}
\hat{\mu}_j(\hat{z}) & = & \frac{1}{\sqrt{2}} \sum_{z \in \mathbb{Z}_2} (-1)^{\hat{z} z} \mu_j(z) \\
  & = & \frac{1}{2} \left[ 1 + (-1)^{\hat{z} \oplus y_j} \cos\theta \right] & \\
  & = &
\begin{cases}
1 - p & \ \text{if}\ \hat{z} = y_j, \\
p     & \ \text{if}\ \hat{z} = y_j \oplus 1,
\end{cases}
\end{IEEEeqnarray}
where $p = \frac{1 - \cos\theta}{2}$.
It follows that
\begin{IEEEeqnarray}{rCl+x*}
\hat{s}(h) & = & \sqrt{2^{n-k}} \hat{s}'(y_h) \\
  & = & 2^{(n-k)/2} \sum_{\hat{x} \in \mathbb{Z}_2^n} 2^{k-n/2} \mathbb{I}(\hat{x} \in \MCC^{\perp}) \prod_{j=1}^n \hat{\mu}_j(\hat{x}_j) \\
  & = & 2^{(n - k + 2k - n)/2} \sum_{\hat{x} \in \mathbb{Z}_2^n} \mathbb{I}(\hat{x} \in \MCC^{\perp}) \ p^{w_H(y_h \oplus \hat{x})} (1 - p)^{n - w_H(y_h \oplus \hat{x})} \\
  & = & 2^{k/2} \sum_{z \in y_h\, \oplus\, \MCC^{\perp}} p^{w_H(z)} (1 - p)^{n - w_H(z)}.
\end{IEEEeqnarray}
This completes the derivation of~\eqref{eq:shat_final}. \hfill \IEEEQEDhere
\end{IEEEproof}


\begin{remark}
For the case where the $i$-th bit is transmitted over a PSC with parameter $\theta_i$, the dual channel for the $i$-th bit is a BSC with error probability $p_i = \frac{1-\cos\theta_i}{2}$.
In that case, the term $p^{w_H (z)} (1-p)^{n-w_H (z)}$ in Lemma~\ref{lem:FT_overlap} changes to
$ \prod_{i=1}^N p_i^{[c_g]_i} (1-p_i)^{1-[c_g]_i}$ and $2^{-k/2}\hat{s}(h)$ has the same interpretation but with the newå error probabilities.
\end{remark}

Now we can use this result to calculate the inner product between SRM measurement vectors and the codeword states.

\begin{lemma}
\label{lem:srm_innerpdt}
Consider an $[n,k]$ binary linear code $\MCC$ and the channel $\psc(\theta)$. 
The overlap between the square root measurement (SRM) vectors and the states obtained by transmitting the codewords of $\MCC$ over $\psc(\theta)$ is given by
\begin{align} \label{eq:srm_phi_overlap}
\left| \dbraket{\psi_g}{\phi_t} \right|^2 & = \frac{ \hat{\sigma}(g \oplus t)^2 }{2^{k}}.
\end{align}
This is equal to the probability of sending a message $t \in \mathbb{Z}_2^k$ and decoding it as $g \in \mathbb{Z}_2^k$ using the SRM.
\end{lemma}
\begin{IEEEproof}
From the expression~\eqref{eq:shat_final} it is evident that $\sigma(h) = 2^{k/4} \sqrt{\hat{s}(h)} \neq 0$ for all $h \in \mathbb{Z}_2^k$.
Hence, using the expression for $\dket{\psi_g}$ in~\eqref{eq:psi_g}, we can compute the inner product between $\dket{\psi_g}$ and $\dket{\phi_{t}} = Z(c_t) \dket{\theta}^{\otimes n}$, for $g,t \in \mathbb{Z}_2^k$, as follows.
\begin{IEEEeqnarray}{rCl+x*}
\dbraket{\psi_g}{\phi_t} & = & \frac{1}{2^k} \sum_{h,f \in \mathbb{Z}_2^k} \frac{(-1)^{h (f \oplus g)^T}}{2^{k/4} \sqrt{\hat{s}(h)}} \dbra{\theta}^{\otimes n} Z(c_f) Z(c_t) \dket{\theta}^{\otimes n} & \\
  & = & \frac{1}{2^{5k/4}} \sum_{h,f \in \mathbb{Z}_2^k} \frac{(-1)^{h (f \oplus g)^T}}{\sqrt{\hat{s}(h)}} \left( \cos\theta \right)^{w_H(c_f \oplus c_t)} & \\
  & = & \frac{1}{2^{5k/4}} \sum_{h \in \mathbb{Z}_2^k} \frac{(-1)^{h g^T}}{\sqrt{\hat{s}(h)}} \left[ \sum_{f \in \mathbb{Z}_2^k} (-1)^{h f^T} \left( \cos\theta \right)^{w_H(c_{f \oplus t})} \right] & \\
  & = & \frac{1}{2^{5k/4}} \sum_{h \in \mathbb{Z}_2^k} \frac{(-1)^{h (g \oplus t)^T}}{\sqrt{\hat{s}(h)}} \left[ \sum_{f' \in \mathbb{Z}_2^k} (-1)^{h (f')^T} \left( \cos\theta \right)^{w_H(c_{f'})} \right] & \\
  & = & \frac{1}{2^{k/4}} \left[ \frac{1}{\sqrt{2^k}} \sum_{h \in \mathbb{Z}_2^k} (-1)^{h (g \oplus t)^T} \sqrt{\hat{s}(h)} \right] & \\
  & = & \frac{1}{2^{k/2}} \left[ \frac{1}{\sqrt{2^k}} \sum_{h \in \mathbb{Z}_2^k} (-1)^{h (g \oplus t)^T} \sigma(h) \right] & \\
  & = & \frac{ \hat{\sigma}(g \oplus t) }{2^{k/2}}.
\end{IEEEeqnarray}
The stated result follows immediately. \hfill \IEEEQEDhere
\end{IEEEproof}

This was also used recently to verify the optimality of the BPQM algorithm for decoding a $5$-bit code over $\psc(\theta)$~\cite{Rengaswamy-arxiv20}.
Finally, we produce a closed-form expression for the optimal probability of (block) error $P_e$.

\begin{theorem}
\label{thm:Pe_PSC}
Given an $[n,k]$ binary linear code $\MCC$, the optimal block error probability for transmission over the channel $\psc(\theta)$ is given by
\begin{align}
\label{eq:Pe_PSC}
P_e & = 1 - \mathcal{B}\left(  \frac{\hat{s}(\cdot)}{2^{k/2}} , \nu(\cdot) \right)^2,
\end{align}
where the Bhattacharyya coefficient between $ 2^{-k/2} \hat{s}(h)$~\eqref{eq:shat_final} and the uniform distribution $\nu (h)= 2^{-k}$ is defined by
\begin{align}
\mathcal{B}\left(  \frac{\hat{s}(\cdot)}{2^{k/2}} , \nu(\cdot) \right) & \coloneqq \sum_{h \in \mathbb{Z}_2^k} \sqrt{\frac{\hat{s}(h)}{2^{k/2}}} \sqrt{\frac{1}{2^k}}.
\end{align}
\end{theorem}
\begin{IEEEproof}
We substitute the inner product $\left| \dbraket{\psi_g}{\phi_t} \right|^2$ from Lemma~\ref{lem:srm_innerpdt} in~\eqref{eq:Pe_reduced} to calculate
\begin{align}
P_e & = \frac{1}{2^k} \sum_{j \in \mathbb{Z}_2^k} \left[ \left( \sum_{i \in \mathbb{Z}_2^k} \left| \dbraket{\psi_j}{\phi_i} \right|^2 \right) - \left| \dbraket{\psi_j}{\phi_j} \right|^2 \right] \\
  & = \frac{1}{2^k} \sum_{j \in \mathbb{Z}_2^k} \left[ \left( \sum_{i \in \mathbb{Z}_2^k} \frac{ \hat{\sigma}(j \oplus i)^2 }{2^{k}} \right) - \frac{ \hat{\sigma}(0)^2 }{2^{k}} \right] \\
  & = \left( \sum_{i \in \mathbb{Z}_2^k} \frac{ \hat{\sigma}(i)^2 }{2^{k}} \right) - \frac{ \hat{\sigma}(0)^2 }{2^{k}} \\
  & \overset{\text{(a)}}{=} \left( \sum_{i \in \mathbb{Z}_2^k} \frac{ \sigma(i)^2 }{2^{k}} \right) - \frac{1}{2^{k}} \left( \frac{1}{\sqrt{2^k}} \sum_{i \in \mathbb{Z}_2^k} \sigma(i) \right)^2 \\
\label{eq:sum_sqrt_shat}
  & = \left( \sum_{i \in \mathbb{Z}_2^k} \frac{ 2^{k/2} \hat{s}(i) }{2^{k}} \right) - \frac{2^{k/2}}{2^{2k}} \left( \sum_{i \in \mathbb{Z}_2^k} \sqrt{\hat{s}(i)} \right)^2 \\
  & = s(0) - \left( \sum_{i \in \mathbb{Z}_2^k} \sqrt{\frac{\hat{s}(i)}{2^{k/2}}} \sqrt{\frac{1}{2^k}} \right)^2 \\
  & = 1 - \mathcal{B}\left(  \frac{\hat{s}(\cdot)}{2^{k/2}} , \nu(\cdot) \right)^2,
\end{align}
where in step (a) we have used Parseval's identity for Fourier transforms.
\end{IEEEproof}

\subsection{Bhattacharyya Coefficient for Secret Communication on the BSC}
\label{sec:wyner_bsc}

In order to interpret the sum in~\eqref{eq:shat_final} for the BSC, and hence $P_e$, consider the secrecy problem using the code $\MCC^{\perp}$, which has $2^k$ cosets in $\mathbb{Z}_2^n$.
We will try to keep the notation consistent with the PSC calculation.
When we want to send the uniform random message $h \in \mathbb{Z}_2^k$, we transmit $y_{h,c} = h G_{\MCC^{\topbot}} \oplus c$ for some randomly chosen codeword $c \in \MCC^{\perp}$.
So, the message is encoded into the coset of $\MCCd$ generated by $h G_{\MCC^{\topbot}}$, and the randomness from $c$ adds uncertainty to protect the message from the eavesdropper.
In~\cite{Renes-it18}, this is referred to as \emph{randomized encoding} into $\MCC^{\topbot}$.
The channel is $n$ independent uses of $\bsc(p)$, so we receive $\hat{x} = y_{h,c} \oplus e$, where $e_i \in \mathbb{Z}_2$ are i.i.d. Bernoulli($p$).
If we calculate the syndrome w.r.t. the parity-check matrix of $\MCC^{\perp}$, then using the observations in~\eqref{eq:H_relations} we get
\begin{align}
\hat{x} H_{\MCC^{\perp}}^T = (h G_{\MCC^{\topbot}} \oplus c \oplus e) G_{\MCC}^T = h \oplus e G_{\MCC}^T.
\end{align}

At the receiver side, let us calculate the posterior distribution for the messages $\hat{h} \in \mathbb{Z}_2^k$ given the received vector $\hat{x} \in \mathbb{Z}_2^n$.
\begin{align}
\pr{\hat{h}\, \big\vert\, \hat{x}} & = \frac{ \pr{\hat{x}\, \big\vert\, \hat{h}} \cdot \pr{\hat{h}} }{ \sum_{\tilde{h} \in \mathbb{Z}_2^k} \pr{\hat{x}\, \big\vert\, \tilde{h}} \cdot \pr{\tilde{h}} } \\
  & = \frac{ \pr{\hat{x}\, \big\vert\, \hat{h}} }{ \sum_{\tilde{h} \in \mathbb{Z}_2^k} \pr{\hat{x}\, \big\vert\, \tilde{h}} } \\
  & = \frac{ \sum_{c \in \MCC^{\perp}} \pr{\hat{x}\, \big\vert\, \hat{h}, c} \cdot \pr{c} }{ \sum_{\tilde{h} \in \mathbb{Z}_2^k} \sum_{c \in \MCC^{\perp}} \pr{\hat{x}\, \big\vert\, \tilde{h}, c} \cdot \pr{c} } \\
  & = \frac{ \sum_{c \in \MCC^{\perp}} \pr{e = (\hat{h} G_{\MCC^{\topbot}} \oplus c) \oplus \hat{x}\, \big\vert\, \hat{h}, c} }{ \sum_{\tilde{h} \in \mathbb{Z}_2^k} \sum_{c \in \MCC^{\perp}} \pr{e = (\tilde{h} G_{\MCC^{\topbot}} \oplus c) \oplus \hat{x}\, \big\vert\, \tilde{h}, c} } \\
  & \overset{\text{(a)}}{=} \frac{ \sum_{c \in \MCC^{\perp}} \pr{e = y_{\hat{h},c} \oplus \hat{x}} }{ 1 }\ ;\ \ \hat{x} \coloneqq \hat{h}' G_{\MCC^{\topbot}} \oplus c'\ \text{for\ some}\ \hat{h}' \in \mathbb{Z}_2^k,\ c' \in \MCC^{\perp} \\
  & = \pr{\text{error} \in (\hat{h} \oplus \hat{h}')\, G_{\MCC^{\topbot}} \oplus\, \MCC^{\perp}},
\end{align}
where step (a) uses the fact that $\mathbb{Z}_2^n$ is a direct sum of $\MCCd$ and $\MCC^{\topbot}$ to express $\hat{x}$ uniquely as $\hat{x} \coloneqq \hat{h}' G_{\MCC^{\topbot}} \oplus c'$ for some $\hat{h}' \in \mathbb{Z}_2^k,\ c' \in \MCC^{\perp}$.
This also explains why the denominator of the equation before step (a) equals 1 (e.g., it is a sum over all error patterns).
Continuing, we see that
\begin{align}
\pr{\hat{h}\, \big\vert\, \hat{x}}  & = \sum_{c \in \MCC^{\perp}} p^{w_H(e)} (1 - p)^{n - w_H(e)}\ ; \ \ e = (\hat{h} \oplus \hat{h}')\, G_{\MCC^{\topbot}} \oplus c \\
  & = \sum_{z \in y_{\hat{h},0}\, \oplus\, \MCC^{\perp}} p^{w_H(z)} (1 - p)^{n - w_H(z)} \ \ \text{if}\ \ \hat{x} \in \MCC^{\perp} (\Leftrightarrow \hat{h}' = 0) \\
  & = \frac{\hat{s}(\hat{h})}{2^{k/2}}.
\end{align}

Hence, we see that the quantity $\hat{s}(h)$ in~\eqref{eq:shat_final} is calculating the posterior for the coset represented by $h$ when the received vector $\hat{x}$ is in $\MCC^{\perp}$, i.e., the zero coset with $\hat{h}' = 0$.
We also observe that, when $\hat{x}$ is in a different coset of $\MCC^{\perp}$, the posterior probabilities are just permuted according to that coset.
Let $\pi_{\hat{x}} \colon \mathbb{Z}_2^k \to \mathbb{Z}_2^k$ denote the implied permutation for $\hat{x}\in\mathbb{Z}_2^n $, which implies that $\pi_{\hat{x}}(\hat{h}) = \hat{h} \oplus \hat{h}'$.
Hence, $\pr{\hat{h}\, \big\vert\, \hat{x}} = 2^{-k/2} \hat{s}(\pi_{\hat{x}}(\hat{h}))$.

However, the measure of secrecy is the squared Bhattacharyya distance between the posterior distribution and the uniform distribution $\nu(h)=2^{-k}$ on all cosets. 
Since this quantity is independent of the mapping of the posterior probabilities to the cosets, the aforementioned permutation is inconsequential.
Specifically, the secrecy measure we consider is given by the squared Bhattacharyya coefficient, 
\begin{align}
\label{eq:Secrecy_BSC}
\mathcal{B}\left(  \frac{\hat{s}(\cdot)}{2^{k/2}} , \nu(\cdot) \right)^2 & = \left(\sum_{h \in \mathbb{Z}_2^k} \sqrt{\frac{\hat{s}(h)}{2^{k/2}}} \sqrt{\frac{1}{2^k}}\right)^2 .
\end{align}

In quantum information, the fidelity between two density matrices $\rho$ and $\sigma$ is defined as $\mathcal{F}(\rho,\sigma) \coloneqq \norm{\sqrt{\rho} \sqrt{\sigma}}_1^2$, where $\norm{M}_1 \coloneqq \tr{\sqrt{M^{\dagger} M}}$ is the trace norm of a matrix $M$.
For classical distributions, we can take $\rho$ and $\sigma$ to be diagonal density matrices with the diagonal elements being the respective probabilities.
For our case, take the diagonal elements of $\rho$ to be $2^{-k/2} \hat{s}(h)$ for $h \in \mathbb{Z}_2^k$ and set $\sigma = \frac{1}{2^k} I_{2^k}$.
Then it is easy to check that 
\begin{align}
\mathcal{F}(\rho,\sigma) = \mathcal{B}\left(  \frac{\hat{s}(\cdot)}{2^{k/2}} , \nu(\cdot) \right)^2 .
\end{align}

To understand this result in terms of channels, let $W_{\MCC}^{\psc(\theta)}$ denote the CQ channel implied by the channel coding problem over PSC($\theta$) using $\MCC$ (i.e., the input alphabet is $\{0,1\}^k$).
Similarly, let $W_{\MCC^{\perp}}^{\bsc(p)}$ denote the CQ channel implied by sending the coset selector $s\in \{0,1\}^k$ for the secrecy problem using $\MCC^\perp$ over the BSC($p$).
Then, comparing~\eqref{eq:Pe_PSC} and~\eqref{eq:Secrecy_BSC}, we have shown a tight duality between channel coding over the PSC and secret communication over the BSC as measured by the Bhattacharyya distance, i.e., fidelity. 
This proves the duality result in~\cite[Corollary 3]{Renes-it18} for the PSC-BSC special case, since $(1 - P_e)$ is exactly the optimal guessing probability for the PSC, $P\big(W_{\MCC}^{\psc(\theta)}\big)$, and the above fidelity is the measure of ``decoupling'' for secrecy over the BSC, $Q\big(W_{\MCC^{\perp}}^{\bsc(p)}\big)$.
However, our methodology allowed us to derive the result using only the square root measurement and the discrete Fourier transform based tool set borrowed from~\cite{Eldar-it00}.

In the next section, we discuss what these results imply for Von Neumann entropy.
After that, in the following two sections, we complete the duality picture by focusing on the block error rate of channel coding on the BSC and the Bhattacharyya decoupling of secret communications on the PSC.

\subsection{Duality Under Von Neumann Entropy}

The derivations in Sections~\ref{sec:cc_psc} and~\ref{sec:wyner_bsc} also reveal the following general result.
Given a binary linear code $\MCC$ for transmission over the $\psc(\theta)$, the Fourier transform $\hat{s}(h)$ of the overlap function $s(g)$ on the channel outputs (with respect to the zero codeword) gives the posterior distribution for secret communications on the dual channel $\bsc(\frac{1 - \cos\theta}{2})$, when using the cosets of $\MCCd$ in Wyner's wire-tap coding scheme.
Moreover, the conditional probabilities of all possible secret messages, $\{2^{-k/2}\hat{s}(h), h\in \{0,1\}^k\}$, appear as the eigenvalues of the density matrix for the PSC observation $Y$ in the channel coding problem, $\rho^{Y,S=0} = 2^{-k} \Phi \Phi^\dagger$.

\begin{lemma}
\label{lem:cc_psc_Gram_eigenvalues}
For the channel coding problem on the PSC, the overlap (or normalized Grammian) matrix $\Gamma = 2^{-k} \Phi^\dagger \Phi$ is diagonalized by the Fourier transform $\mathcal{F}$. From this, we find that the set of (non-zero) eigenvalues of both $\Gamma$ and the density matrix $\rho^{Y,S=0}$, equal the set of non-zero elements in $\{2^{-k/2}\hat{s}(h) \,|\, h\in \{0,1\}^k\}$.
\end{lemma}
\begin{IEEEproof}
We can express $\mathcal{F}$ and $\Phi^\dagger \Phi$ as follows:
\begin{align}
\mathcal{F} = \frac{1}{\sqrt{2^k}} \sum_{h,g \in \mathbb{Z}_2^k} (-1)^{hg^T} \dketbra{h}{g} \quad , \quad 
\Phi^\dagger \Phi = \sum_{h,g \in \mathbb{Z}_2^k} \left( \cos\theta \right)^{w_H(c_h \oplus c_g)} \dketbra{h}{g} = \sum_{h,g \in \mathbb{Z}_2^k} \left( \cos\theta \right)^{w_H(c_{h \oplus g})} \dketbra{h}{g}.
\end{align}
Then we observe that
\begin{align}
\mathcal{F} \Phi^\dagger \Phi \mathcal{F}^\dagger & = \frac{1}{2^k} \sum_{h,g \in \mathbb{Z}_2^k} \sum_{h_1,g_1 \in \mathbb{Z}_2^k} \sum_{h_2,g_2 \in \mathbb{Z}_2^k} (-1)^{hg^T + h_2 g_2^T} \left( \cos\theta \right)^{w_H(c_{h_1 \oplus g_1})} \dket{h} \dbraket{g}{h_1} \dbraket{g_1}{h_2} \dbra{g_2} \\
  & = \frac{1}{2^k} \sum_{h,g \in \mathbb{Z}_2^k} \sum_{h_2,g_2 \in \mathbb{Z}_2^k} (-1)^{hg^T + h_2 g_2^T} \left( \cos\theta \right)^{w_H(c_{g \oplus h_2})} \dketbra{h}{g_2} \\
  & = \frac{1}{2^k} \sum_{h,g_2 \in \mathbb{Z}_2^k} \dketbra{h}{g_2} \sum_{g, m \in \mathbb{Z}_2^k} (-1)^{hg^T + (g \oplus m) g_2^T} \left( \cos\theta \right)^{w_H(c_m)} \\
  & = \frac{1}{2^k} \sum_{h,g_2 \in \mathbb{Z}_2^k} \dketbra{h}{g_2} \sum_{m \in \mathbb{Z}_2^k} (-1)^{m g_2^T} \left( \cos\theta \right)^{w_H(c_m)} \left( \sum_{g \in \mathbb{Z}_2^k} (-1)^{g (h \oplus g_2)^T} \right) \\
  & = \frac{1}{2^k} \sum_{h , g_2 \in \mathbb{Z}_2^k} \dketbra{h}{g_2} \sum_{m \in \mathbb{Z}_2^k} (-1)^{m g_2^T} \left( \cos\theta \right)^{w_H(c_m)} \, \left( 2^k \; \mathbb{I}(h=g_2) \right) \\
  & = \sum_{h \in \mathbb{Z}_2^k} \left( \sum_{m \in \mathbb{Z}_2^k} (-1)^{m h^T} \left( \cos\theta \right)^{w_H(c_m)} \right) \dketbra{h} \\
  & = \sum_{h \in \mathbb{Z}_2^k} 2^{k/2} \hat{s}(h) \dketbra{h}.
\end{align}
Thus, the set of eigenvalues of $\Gamma$ equals the set $2^{-k} 2^{k/2} \hat{s}(h) = 2^{-k/2} \hat{s}(h)$.
By computing the SVD, one can show that the set of non-zero eigenvalues of $\Phi^\dagger \Phi$ equals the set of non-zero eigenvalues of $\Phi \Phi^\dagger$.
Thus, the set of non-zero eigenvalues of the observation density matrix $\rho^{Y,S=0} = 2^{-k} \Phi \Phi^\dagger$ also equals the set of non-zero eignvalues of $\Gamma$.
\end{IEEEproof}

Thus, we can also recover the analogous duality result in~\cite{Renes-it18} for the Von Neumann entropy.
For a binary linear code $\MCC$, let $\rho^{Y,S=0} = 2^{-k} \Phi \Phi^\dagger$ be the density matrix for the quantum output $Y$ of the channel coding problem over $\psc(\theta)$ when the coset shift $S=0$.
Since the Von Neumann entropy $H(C)_{\rho}$ of a quantum system $C$ with density matrix $\rho^C$ equals the Shannon entropy of the eigenvalues of $\rho^C$~\cite{Wilde-2013}, it follows that
\begin{align}
H(Y|S=0)_{\rho^{Y,S=0}}
&= \sum_{h\in\{0,1\}^k} 2^{-k/2}\hat{s}(h) \log \frac{1}{2^{-k/2}\hat{s}(h)} \\
&= \sum_{h\in\{0,1\}^k} \pr{\hat{h}\, \vert\, \hat{x}} \log \frac{1}{\pr{\hat{h}\, \vert\, \hat{x}}} \\
&= H(S'|Y'), \label{eq:vne_y_duality}
\end{align} 
where the final quantity is the classical Shannon entropy.
Of course, this result also extends to any other classical entropies of the form $H'({p_1,\ldots,p_M}) = \sum_{m=1}^M f(p_m)$ (e.g., R\'{e}nyi entropy) and their natural quantum analogues.

Using the same setup as the BEC analysis in Section~\ref{sec:cc_sec_duality_bec}, we can use the above result to investigate the Von Neumann conditional entropy.
Similar to the BEC result in~\eqref{eq:dualsecrecy_vs_primalcoding}, this implies that
\begin{align} 
H(U|Y,S=0)_{\rho^{UY,S=0}}
&= H(U|S=0)+H(Y|U,S=0)_{\rho^{UY,S=0}}-H(Y|S=0)_{\rho^{Y,S=0}} \\
&= k+0-H(S'|Y'). \label{eq:neumann_duality}
\end{align} 
Next, consider BEC channel coding duality result defined by~\eqref{eq:entropy_duality}.
One can generalize this to the PSC by observing
\begin{align}
H(U'|Y',S')
&= H(U'|S') + H(Y'|S',U') - H(Y'|S') \\
&= n-k + \eta(p) \, n - \big( H(Y')+H(S'|Y')-H(S') \big) \\
&= n-k + \eta(p) \, n - \left( n + \big(k-H(U|Y,S=0)_{\rho^{UY,S=0}} \big) -k \right) \\
&= H(U|Y,S=0)_{\rho^{UY,S=0}} + \eta(p) \, n - k, \label{eq:vne_coding_duality}
\end{align}
where $\eta(p) \triangleq -p \log p - (1-p) \log (1-p)$ is the binary entropy function and the term $\eta(p)\,n$ can be seen as the total entropy produced by the dual channel.


\vspace{2mm}
\noindent \textbf{GEXIT Function Duality}
\vspace{2mm}

The form of~\eqref{eq:vne_coding_duality} also allows one to derive a GEXIT duality formula via differentiation.
To see this, we will again use the setup from the BEC analysis in Section~\ref{sec:cc_sec_duality_bec}.
For notational convenience, we introduce the random variable $X' \triangleq X$ and use it to indicate when we are interpreting $X$ as the transmitted vector for the dual system.
We will begin by adjusting the setup and rewriting~\eqref{eq:vne_coding_duality} with two changes.
First, we note that $H(U'|Y',S') = H(X'|Y',S')$ because $X' = S' E + U' H$ and $H(U|Y,S=0)_{\rho^{UY,S=0}} = H(X|Y,S=0)_{\rho^{UY,S=0}}$ because $X = U G + S F$.
Next, we assume that $Y_i$ is an observation of $X_i$ through a PSC with parameter $\theta_i$.
Using duality, this implies that $Y_i '$ should be an observation of $X_i '$ through a BSC with error probability $p_i = \frac{1-\cos \theta_i}{2}$.
Using these modifications, we can rewrite~\eqref{eq:vne_coding_duality} as
\begin{equation} \label{eq:gexit_setup}
H(X'|Y'(h_1 ',\ldots,h_n '),S') = H(X|Y(h_1,\ldots,h_n),S=0)_{\rho^{UY,S=0}} + \sum_{i=1}^n h_i' - k,
\end{equation}
where $h_i' = \eta(p_i) = H(X_i'|Y_i')$ is the input entropy for a BSC observation with error probability $p_i$, $h_i=1-h_i' = H(X_i | Y_i)$ is the input entropy for a PSC observation with parameter $\theta_i = \cos^{-1} (1-2p_i)$, and the dependence of $Y'$ and $Y$ on these entropies is shown explicitly.

Using the definition in~\cite[Def.~4.152]{RU-2008}, we see that the GEXIT function for the $i$-th bit of $\MCCd$ on the BSC is given by
\begin{align}
g_i ' (h_1',\ldots,h_n')
&= \frac{\mathrm{d}}{\mathrm{d} h_i'} H(X'|Y'(h_1',\ldots,h_n'),S') \\
&= \frac{\mathrm{d}}{\mathrm{d} h_i'} \big(H(X_i '|Y'(h_1',\ldots,h_n'),S') + H(X_{\sim i} '|X_i',Y'(h_1',\ldots,h_n'),S') \big) \\
&= \frac{\mathrm{d}}{\mathrm{d} h_i'} H(X_i '|Y'(h_1',\ldots,h_n'),S'), \label{eq:gexit_dual}
\end{align}
because $H(X_{\sim i} '|X_i',Y'(h_1',\ldots,h_n'),S')$ does not depend on $h_i '$. Using the same idea, we define the GEXIT function for the $i$-th bit of $\MCC$ on the PSC to be
\begin{align}
g_i (h_1,\ldots,h_n)
&= \frac{\mathrm{d}}{\mathrm{d} h_i} H(X|Y(h_1,\ldots,h_n),S=0)_{\rho^{UY,S=0}} \\
&= \frac{\mathrm{d}}{\mathrm{d} h_i} H(X_i|Y(h_1,\ldots,h_n),S=0)_{\rho^{UY,S=0}}. \label{eq:gexit_primal}
\end{align}
Since $h_i ' = 1-h_i$, we can combine \eqref{eq:gexit_setup}, \eqref{eq:gexit_dual}, and \eqref{eq:gexit_primal} to see that
\begin{equation} \label{eq:gexit_duality}
g_i ' (h_1',\ldots,h_n') = 1 - g_i  (h_1,\ldots,h_n).
\end{equation}
Thus, the BSC GEXIT curve exactly satisfies the expected duality formula in terms of the PSC GEXIT function.
This can also be extended to general BMS channels and heralded mixtures of PSCs by averaging over the implied parameter distributions.

Perhaps, it provides a slight extension of the EXIT function duality relationship for $\MCC$ and $\MCCd$~\cite[Eqn.~74]{Renes-it18}, when restricted to Von Neumann entropy and the pure-state channel. This is because~\eqref{eq:gexit_duality} with $h_i = 1$ and $h_i' = 0$ reduces to the EXIT function duality relationship
$$ H(X_i ' | Y_{\sim i}',S') = 1 - H(X_i | Y_{\sim i},S=0)_{\rho^{X_i Y,S=0}}.$$
Showing this reduction for $g_i ' (h_1',\ldots,h_n')$ only requires a standard GEXIT calculation.
But, for $g_i  (h_1,\ldots,h_n)$, the calculation would lead us too far from the main point of this paper.
Instead, Appendix~\ref{sec:exit_appendix} contains an straightforward derivation of this final statement.

\subsection{Channel Coding on the BSC}

We will now use $\MCC^{\perp}$ to perform standard channel coding over the BSC.
However, we will setup this problem as a state discrimination, or hypothesis testing, problem with the BSC as a ``classical-quantum'' channel.
Therefore, given a message $z \in \mathbb{Z}_2^{n-k}$, we transmit the codeword $c_z = z G_{\MCC^{\perp}} = z H_{\MCC}$ and receive $y = c_z \oplus E$ for some random error $E \in \mathbb{Z}_2^n$ that is i.i.d. Bernoulli($p$) with $p = (1 - \cos\theta)/2$.
This specific choice of $p$ is chosen in order to tie this setup to secrecy on $\psc(\theta)$ in the next section.
To set this up as a CQ state discrimination problem, the receiver defines the density matrices
\begin{align}
\label{eq:cc_bsc_density_matrices}
\varphi_z \coloneqq \sum_{e \in \mathbb{Z}_2^n} P_E (e) 
\dketbra{c_z \oplus e} = \sum_{e \in \mathbb{Z}_2^n} p^{w_H(e)} (1-p)^{n - w_H(e)} \dketbra{c_z \oplus e}
\end{align}
corresponding to each message $z$.
Observe that $\dket{c_z \oplus e} \in \mathbb{C}^{2^n}$ are all standard basis vectors. 
So, each $\varphi_z$ is a diagonal matrix (indicating that it is essentially classical) with entries $P_E (e)$ 
permuted by $c_z$.
The matrix $\varphi_z$ can be interpreted as a discrete conditional probability distribution on all possible received vectors, given that $c_z$ was transmitted.
Note that the eigenvalues of the average output state $\frac{1}{2^{n-k}} \sum_{c_z \in \MCCd} \varphi_z$ can be expressed using $\hat{s}(h)$~\eqref{eq:shat_final}.
In the language of~\cite{Renes-it18}, the CQ state (density matrix) relevant to this state discrimination problem is given by
\begin{align}
\label{eq:coding_bsc_cq_state}
\Psi_{\hat{A} \bar{A} C^n D^n} = \frac{1}{2^{n-k}} \sum_{z \in \mathbb{Z}_2^{n-k}} \dketbra{z}_{\hat{A}} \otimes \dketbra{0}_{\bar{A}} \otimes (\varphi_z)_{C^n D^n}.
\end{align}
Here, subsystem $\hat{A}$ (resp. $\bar{A}$) corresponds to the message space of $\MCCd$ (resp. $\MCC^{\topbot}$), and $C^n D^n$ together hold the output density matrix $\varphi_z$ for the message in $\hat{A}$ (see~\cite{Renes-it18} for more details).
Intuitively, the above state represents a joint distribution between the transmitted message and received state.
Hence, given the candidate states $\{ \varphi_z , \ z \in \mathbb{Z}_2^{n-k} \}$, we need to determine the POVM that optimally distinguishes them, i.e., the POVM that effectively induces the classical MAP decoder in this CQ setup.

Recollect that the optimal block MAP success probability is given by
\begin{align}
\mathbb{P}\left[ \text{MAP success for $\MCC^{\perp}$ on BSC$(p)$} \right] = \sum_{m \in \mathbb{Z}_2^{k}} \max_{ u \in (m G_{\MCC^{\topbot}} \oplus \MCC^{\perp}) } p^{w_H(u)} (1-p)^{n - w_H(u)}.
\end{align}
It is well-known that, on the BSC, MAP decoding can be implemented by determining the coset of the received vector w.r.t. cosets of $\MCCd$, and then choosing the corresponding coset leader as the error introduced by the channel.
The coset leaders are the vectors of minimum weight in the respective cosets.
Hence, the above expression simply calculates the probability that the error vector is one of the coset leaders.

We use the shorthand $p_e = P_E (e)$ 
throughout the remainder of the paper.

\begin{definition}
\label{def:min_wt_vec}
For a given $v \in \mathbb{Z}_2^n$, let $$v^* \triangleq \argmax_{u \in v \oplus \MCC^{\perp}} p_u = \argmax_{u \in v \oplus \MCC^{\perp}} p^{w_H(u)} (1 - p)^{n - w_H(u)}$$ be an arbitrary minimum-weight vector in the coset $v\, \oplus\, \MCC^{\perp}$ (e.g., ties can broken broken via lexicographical ordering).
Using this, we treat the decorator mapping $(\cdot)^*$ as a function mapping $\mathbb{Z}_2^n$ to itself defined by $v \mapsto v^*$.
\end{definition}

\begin{theorem}
\label{thm:BSC_MAP}
The MAP decoder for channel coding on the $\bsc(p)$ with the $[n,n-k]$ binary linear code $\MCCd$ can be implemented using the projective measurement defined by
\begin{align}
\left\{ \Pi_z \coloneqq \sum_{v \in \MCC^{\topbot}} \dketbra{c_z \oplus v^*} ; \ z \in \mathbb{Z}_2^{n-k} \right\}.
\end{align}
\end{theorem}
\begin{IEEEproof}
The success probability for this scheme can be calculated as
\begin{align}
\mathbb{P}\left[ \text{Success} \right] & \coloneqq \sum_{z \in \mathbb{Z}_2^{n-k}} \frac{1}{2^{n-k}} \tr{ \varphi_z \cdot \Pi_z } \ \ (\text{using~\eqref{eq:density_msmt_prob}}) \\
  & = \frac{1}{2^{n-k}} \sum_{z \in \mathbb{Z}_2^{n-k}} \tr{ \sum_{e \in \mathbb{Z}_2^n} p_e \dketbra{c_z \oplus e} \cdot \sum_{v \in \MCC^{\topbot}} \dketbra{c_z \oplus v^*} } \\
  & = \frac{1}{2^{n-k}} \sum_{z \in \mathbb{Z}_2^{n-k}} \tr{ \sum_{v' \in \mathbb{Z}_2^n} p_{c_z \oplus v'} \dketbra{v'} \cdot \sum_{v \in \MCC^{\topbot}} \dketbra{c_z \oplus v^*} } \\
  & = \frac{1}{2^{n-k}} \sum_{z \in \mathbb{Z}_2^{n-k}} \tr{ \sum_{v \in \MCC^{\topbot}} p_{c_z \oplus (c_z \oplus v^*)} \dketbra{c_z \oplus v^*} } \\
  & = \frac{1}{2^{n-k}} \sum_{z \in \mathbb{Z}_2^{n-k}} \sum_{v \in \MCC^{\topbot}} p_{v^*} \\
  & = \sum_{v \in \MCC^{\topbot}} p_{v^*} \\
  & = \sum_{m \in \mathbb{Z}_2^{k}} \max_{ u \in (m G_{\MCC^{\topbot}} \oplus \MCC^{\perp}) } p^{w_H(u)} (1-p)^{n - w_H(u)} \\
  & = \mathbb{P}\left[ \text{MAP success for $\MCC^{\perp}$ on BSC$(p)$} \right].
\end{align}
Since $\mathbb{Z}_2^n$ is a direct sum of $\MCCd$ and $\MCC^{\topbot}$, it is clear that $\sum_{z \in \mathbb{Z}_2^{n-k}} \Pi_z = I_{2^n}$ and $\Pi_z \Pi_{z'} = \delta_{zz'} \Pi_z$.
Hence, $\left\{ \Pi_z ; \ z \in \mathbb{Z}_2^{n-k} \right\}$ is the optimal POVM corresponding to the MAP decoder.
\end{IEEEproof}

Note that the projector $\Pi_z$ corresponding to the message $z$ simply checks if one of the coset leaders got added to the candidate codeword $c_z$ during transmission over the channel. \\

\noindent \textbf{Square Root Measurement (SRM) is Suboptimal for Decoding on the BSC} \\

Recall that for channel coding over the PSC, we used the geometrically uniform state set criterion for pure states to conclude that the SRM was optimal.
This criterion has subsequently been generalized to mixed states in~\cite{Eldar-it04}.
According to their definition, the hypothesis states $\{ \varphi_z ; \ z \in \mathbb{Z}_2^{n-k} \}$ form a geometrically uniform state set with the relevant group being $\mathcal{G} = \{ X(c_z) ;\ z \in \mathbb{Z}_2^{n-k} \}$, i.e., $\varphi_z = X(c_z)\, \varphi\, X(c_z)$ with $\varphi \coloneqq \varphi_0$.
Here, $X(c_z)$ is defined in an analogous manner to $Z(c_z)$ in Section~\ref{sec:psc}, by simply replacing $Z$ with $X$.
So, using the results of~\cite{Eldar-it04}, we proceed to calculate the SRM for this problem to see if it satisfies their sufficient condition for optimality.

First, since the candidate states are diagonal in the computational basis, we realize that their \emph{factors} (as defined in~\cite{Eldar-it04}) are
\begin{align}
\phi_z = \sum_{e \in \mathbb{Z}_2^n} \sqrt{p_e} \dketbra{c_z \oplus e},
\end{align}
i.e., $\varphi_z = \phi_z \phi_z^{\dagger}$.
Given the uniform prior assumption on the message $z$, the SRM, also called the least squares measurement (LSM) or the pretty good measurement (PGM), in this scenario is given by the POVM $\{ \Pi_z = \mu_z \mu_z^{\dagger} ; \ z \in \mathbb{Z}_2^{n-k} \}$, where 
\begin{align}
\mu_z & \coloneqq \left( \Phi \Phi^{\dagger} \right)^{-1/2} \phi_z, \\
\Phi & \coloneqq 
\left[
\begin{array}{c|c|c|c|c|c}
\phi_{00\cdots 0} & \phi_{00\cdots 1} & \cdots & \phi_z & \cdots & \phi_{11\cdots 1}
\end{array} 
\right].
\end{align}
Then it is clear that $\Phi \Phi^{\dagger}$ is still diagonal in the standard basis and, in particular,
\begin{align}
\Phi \Phi^{\dagger} & = \sum_{z \in \mathbb{Z}_2^{n-k}} \sum_{e \in \mathbb{Z}_2^n} p_e \dketbra{c_z \oplus e} \\
  & = \sum_{z \in \mathbb{Z}_2^{n-k}} \sum_{v \in \mathbb{Z}_2^n} p_{c_z \oplus v} \dketbra{v} \\
  & = \sum_{v \in \mathbb{Z}_2^n} \left( \sum_{c \in \MCC^{\perp}} p_{c \oplus v} \right) \dketbra{v} \\
  & \eqqcolon \sum_{v \in \mathbb{Z}_2^n} \alpha'(v) \dketbra{v}, \\
\alpha'(v) & \coloneqq \sum_{c \in \MCC^{\perp}} p_{c \oplus v} = \sum_{c \in \MCC^{\perp}} p^{w_H(v \oplus c)} (1-p)^{n - w_H(v \oplus c)}.
\end{align}
To check the sufficiency condition for SRM optimality in~\cite{Eldar-it04}, we calculate
\begin{align}
\phi^{\dagger} \left( \Phi \Phi^{\dagger} \right)^{-1/2} \phi & = \sum_{e \in \mathbb{Z}_2^n} \sqrt{p_e} \dketbra{e} \cdot \sum_{v \in \mathbb{Z}_2^n} \frac{1}{\sqrt{\alpha'(v)}} \dketbra{v} \cdot \sum_{e' \in \mathbb{Z}_2^n} \sqrt{p_{e'}} \dketbra{e'} \\
  & = \sum_{v \in \mathbb{Z}_2^n} \frac{p_v}{\sqrt{\alpha'(v)}} \dketbra{v} \\
  & = \sum_{v \in \mathbb{Z}_2^n} \dfrac{ p^{w_H(v)} (1-p)^{n - w_H(v)} }{ \sqrt{ \sum_{c \in \MCC^{\perp}} p^{w_H(v) + w_H(c) - 2 vc^T}  (1-p)^{n - w_H(v) - w_H(c) + 2 vc^T} } } \dketbra{v} \\
  & = \sum_{v \in \mathbb{Z}_2^n} \left[ \sum_{c \in \MCC^{\perp}} p^{-w_H(v) + w_H(c) - 2 vc^T}  (1-p)^{-n + w_H(v) - w_H(c) + 2 vc^T} \right]^{-1/2} \dketbra{v} \\
  & = (1-p)^{n/2} \sum_{v \in \mathbb{Z}_2^n} \left[ \sum_{c \in \MCC^{\perp}} \left( \frac{p}{1-p} \right)^{w_H(c)} \left( \frac{1-p}{p} \right)^{w_H(v) + 2vc^T} \right]^{-1/2} \dketbra{v} \\
  & = (1-p)^{n/2} \sum_{v \in \mathbb{Z}_2^n} \left( \frac{p}{1-p} \right)^{w_H(v)/2} \left[ \sum_{c \in \MCC^{\perp}} \left( \frac{p}{1-p} \right)^{w_H(c)} \left( \frac{1-p}{p} \right)^{2vc^T} \right]^{-1/2} \dketbra{v}.
\end{align}
Hence, in general, this is not a scalar multiple of the identity matrix as required in the sufficient condition. 
So, we cannot conclude that the SRM is optimal.
A sufficient condition for it to be a scalar multiple of the identity is that, for all $v \in \mathbb{Z}_2^n$,
\begin{align}
\sum_{c \in \MCC^{\perp}} \left( \frac{p}{1-p} \right)^{w_H(c)} \left( \frac{1-p}{p} \right)^{2vc^T} & \propto \left( \frac{p}{1-p} \right)^{w_H(v)}.
\end{align}
Nevertheless, we proceed and calculate the probability of block success in this case.

\begin{theorem}
\label{thm:BSC_SRM}
The SRM success probability for channel coding on $\bsc(p)$ with the $[n,n-k]$ binary linear code $\MCCd$ is
\begin{align}
\mathbb{P}\left[ \text{SRM success for $\MCC^{\perp}$ on BSC$(p)$} \right] = \sum_{m \in \mathbb{Z}_2^k} \dfrac{\sum_{u \in (m G_{\MCC^{\topbot}} \oplus \MCC^{\perp})} \left( p^{w_H(u)} (1-p)^{n - w_H(u)} \right)^2}{ \sum_{u \in (m G_{\MCC^{\topbot}} \oplus \MCC^{\perp})} p^{w_H(u)} (1-p)^{n - w_H(u)} }.
\end{align}
This implies that SRM does not achieve the MAP decoder's performance for this problem.
\end{theorem}
\begin{IEEEproof}
Recall that the POVM for the SRM in this case is $\{ \Pi_z = \mu_z \mu_z^{\dagger} ; \ z \in \mathbb{Z}_2^{n-k} \}$, where 
\begin{align}
\mu_z & \coloneqq \left( \Phi \Phi^{\dagger} \right)^{-1/2} \phi_z \\
  & = \sum_{v \in \mathbb{Z}_2^n} \frac{1}{\sqrt{\alpha'(v)}} \dketbra{v} \cdot \sum_{e \in \mathbb{Z}_2^n} \sqrt{p_e} \dketbra{c_z \oplus e} \\
  & = \sum_{v \in \mathbb{Z}_2^n} \frac{1}{\sqrt{\alpha'(v)}} \dketbra{v} \cdot \sum_{v' \in \mathbb{Z}_2^n} \sqrt{p_{c_z \oplus v'}} \dketbra{v'} \\
  & = \sum_{v \in \mathbb{Z}_2^n} \sqrt{\frac{p_{c_z \oplus v}}{\alpha'(v)}} \dketbra{v} \\
\Rightarrow \Pi_z & = \sum_{v \in \mathbb{Z}_2^n} \frac{p_{c_z \oplus v}}{\alpha'(v)} \dketbra{v}.
\end{align}
The probability of block success for the SRM can be calculated as
\begin{IEEEeqnarray}{rCl+x*}
\mathbb{P}\left[ \text{SRM\ Success} \right] & \coloneqq & \sum_{z \in \mathbb{Z}_2^{n-k}} \frac{1}{2^{n-k}} \tr{ \varphi_z \cdot \Pi_z } \ \ (\text{using~\eqref{eq:density_msmt_prob}}) \\
  & = & \frac{1}{2^{n-k}} \sum_{z \in \mathbb{Z}_2^{n-k}} \tr{ \sum_{e \in \mathbb{Z}_2^n} p_e \dketbra{c_z \oplus e} \cdot \sum_{v \in \mathbb{Z}_2^n} \frac{p_{c_z \oplus v}}{\alpha'(v)} \dketbra{v} } \\
  & = & \frac{1}{2^{n-k}} \sum_{z \in \mathbb{Z}_2^{n-k}} \tr{ \sum_{v' \in \mathbb{Z}_2^n} p_{c_z \oplus v'} \dketbra{v'} \cdot \sum_{v \in \mathbb{Z}_2^n} \frac{p_{c_z \oplus v}}{\alpha'(v)} \dketbra{v} } \\
  & = & \frac{1}{2^{n-k}} \sum_{z \in \mathbb{Z}_2^{n-k}} \sum_{v \in \mathbb{Z}_2^n} \frac{p_{c_z \oplus v}^2}{\alpha'(v)} \\
  & = & \frac{1}{2^{n-k}} \sum_{v \in \mathbb{Z}_2^n} \frac{\sum_{c \in \MCC^{\perp}} p_{c \oplus v}^2}{\sum_{c \in \MCC^{\perp}} p_{c \oplus v}} \\
  & = & \frac{1}{2^{n-k}} \sum_{v_a \in \MCC^{\topbot}} \sum_{v_b \in \MCC^{\perp}} \frac{\sum_{c \in \MCC^{\perp}} p_{c \oplus (v_a \oplus v_b)}^2}{\sum_{c \in \MCC^{\perp}} p_{c \oplus (v_a \oplus v_b)}} \\
  & = & \frac{1}{2^{n-k}} \sum_{v \in \MCC^{\topbot}} 2^{n-k} \cdot \frac{\sum_{c \in \MCC^{\perp}} p_{c \oplus v}^2}{\sum_{c \in \MCC^{\perp}} p_{c \oplus v}} \\
  & = & \sum_{v \in \MCC^{\topbot}} \dfrac{\sum_{u \in v \oplus \MCC^{\perp}} \left( p^{w_H(u)} (1-p)^{n - w_H(u)} \right)^2}{\sum_{u \in v \oplus \MCC^{\perp}} p^{w_H(u)} (1-p)^{n - w_H(u)} } \\
  & = & \sum_{m \in \mathbb{Z}_2^k} \dfrac{\sum_{u \in (m G_{\MCC^{\topbot}} \oplus \MCC^{\perp})} \left( p^{w_H(u)} (1-p)^{n - w_H(u)} \right)^2}{ \sum_{u \in (m G_{\MCC^{\topbot}} \oplus \MCC^{\perp})} p^{w_H(u)} (1-p)^{n - w_H(u)} }. & \nonumber \IEEEQEDhere
\end{IEEEeqnarray}
\end{IEEEproof}

\begin{remark}
Note that this success probability can be interpreted in terms of the $1$- and $2$-norms of the length $2^{n-k}$ vector with entries $\{ p^{w_H(u)} (1-p)^{n - w_H(u)} ;\ u \in (m G_{\MCC^{\topbot}} \oplus \MCC^{\perp}) \}$.
Alternatively,  define a random variable $X$ that takes values $\{ x_m ; \ m \in \mathbb{Z}_2^k \}$ with resp. probabilities $\{ q_m \}$, where
\begin{align}
x_m & \coloneqq \dfrac{ \sqrt{ \sum_{u \in (m G_{\MCC^{\topbot}} \oplus \MCC^{\perp})} \left( p^{w_H(u)} (1-p)^{n - w_H(u)} \right)^2 } }{ \sum_{u \in (m G_{\MCC^{\topbot}} \oplus \MCC^{\perp})} p^{w_H(u)} (1-p)^{n - w_H(u)} }, \\
q_m & \coloneqq \sum_{u \in (m G_{\MCC^{\topbot}} \oplus \MCC^{\perp})} p^{w_H(u)} (1-p)^{n - w_H(u)}.
\end{align}
Then we see that the block success probability of SRM is exactly $\mathbb{E}[X^2]$.
\end{remark}

\subsection{Secret Communications on the PSC}

Now we consider using the $[n,k]$ code $\MCC$ for secrecy over the PSC.
More precisely, we have a secret $(n-k)$-bit message to be communicated over a noiseless channel to an intended recipient.
At the same, there is an eavesdropper who observes the input over a pure-state channel with parameter $\theta$.
Similar to the secrecy problem on the BSC, if $h \in \mathbb{Z}_2^{n-k}$ is the message then we transmit the vector $y_{h,c} = h G_{\MCC^{\top}} \oplus c$ for some randomly chosen codeword $c \in \MCC$.
In the language of~\cite{Renes-it18}, this is referred to as \emph{randomized encoding} into $\MCC^{\top}$.
Each bit $y_i$ of $y = y_{h,c}$ is sent over $\psc(\theta)$ to receive $\dket{(-1)^{y_i} \theta} = \cos\frac{\theta}{2} \dket{0} + (-1)^{y_i} \sin\frac{\theta}{2} \dket{1} = Z(y_i) \dket{\theta}$.
But, from the receiver perspective, since there is no information gained until any of the received qubits are measured, the correct representation is that, with equal probability $1/2^{n-k}$, one of the density matrices $\rho_h$ is received, where
\begin{align}
\rho_h & \coloneqq \frac{1}{|\MCC|} \sum_{c \in \MCC} Z(y_{h,c}) \dketbra{\theta}^{\otimes n} Z(y_{h,c}) \\
  & = Z(h G_{\MCC^{\top}}) \left[ \frac{1}{|\MCC|} \sum_{c \in \MCC} Z(c) \dketbra{\theta}^{\otimes n} Z(c) \right] Z(h G_{\MCC^{\top}}) \\
  & \eqqcolon Z(y_h) \left[ \frac{1}{|\MCC|} \sum_{c \in \MCC} Z(c) \dketbra{\theta}^{\otimes n} Z(c) \right] Z(y_h) \\
\Rightarrow \rho_{h'} & = Z(y_h \oplus y_{h'}) \ \rho_h \ Z(y_h \oplus y_{h'}).
\end{align}
Hence, for the resulting hypothesis testing problem, the candidate states $\{ \rho_h = Z(y_h)\, \rho\, Z(y_h) ; \ h \in \mathbb{Z}_2^{n-k} \}$ satisfy the geometrically uniform property as defined for general mixed states in~\cite{Eldar-it04}, where 
\begin{align}
\rho \coloneqq \rho_0 = \frac{1}{|\MCC|} \sum_{c \in \MCC} Z(c) \dketbra{\theta}^{\otimes n} Z(c).
\end{align}

\begin{lemma}
The generator state $\rho$ is diagonalized as
\begin{align}
\rho & = \sum_{m \in \mathbb{Z}_2^k} \lambda(m) \dketbra{\psi(m)}, \\
\label{eq:psi_mhat}
\text{where} \quad \dket{\psi(m)} & = \sum_{v \in (m G_{\MCC^{\topbot}} \oplus \MCC^{\perp})} \frac{ \sqrt{ p^{w_H(v)} \left( 1 - p \right)^{n - w_H(v)} } }{ \sqrt{ \sum_{y \in ( m G_{\MCC^{\topbot}} \oplus \MCC^{\perp} )} p^{w_H(y)} \left( 1-p \right)^{n - w_H(y)} } } \dket{v}, \\
\label{eq:psi_mhat_eval}
\lambda(m) & = \left( \sum_{v \in (m G_{\MCC^{\topbot}} \oplus \MCC^{\perp})} p^{w_H(v)} \left( 1 - p \right)^{n - w_H(v)} \right),
\end{align}
and $p \coloneqq \frac{1 - \cos\theta}{2} = \sin^2\frac{\theta}{2}$ is the same definition that defines the channel parameter for the BSC which is dual to $\psc(\theta)$.
\end{lemma}
\begin{IEEEproof}
Let us start by observing how $\rho$ acts on an arbitrary state $\dket{\psi} = \sum_{v \in \mathbb{Z}_2^n} \psi_v \dket{v}$.
\begin{align}
\rho \dket{\psi} & = \frac{1}{|\MCC|}  \sum_{c \in \MCC} Z(c) \dketbra{\theta}^{\otimes n} Z(c) \cdot \sum_{v \in \mathbb{Z}_2^n} \psi_v \dket{v} \\
  & = \frac{1}{|\MCC|}  \sum_{c \in \MCC} \sum_{v \in \mathbb{Z}_2^n} \psi_v Z(c) \dketbra{\theta}^{\otimes n} (-1)^{cv^T} \dket{v}\ \ (\text{see\ Section~\ref{sec:psc}}) \\
  & = \frac{1}{|\MCC|}  \left[ \sum_{c \in \MCC} \left( \sum_{v \in \mathbb{Z}_2^n} \psi_v (-1)^{cv^T} \left( \sin\frac{\theta}{2} \right)^{w_H(v)} \left( \cos\frac{\theta}{2} \right)^{n - w_H(v)} \right) Z(c) \right] \dket{\theta}^{\otimes n} \\
  & = \frac{1}{|\MCC|}  \left[  \sum_{v \in \mathbb{Z}_2^n} \psi_v \left( \sin\frac{\theta}{2} \right)^{w_H(v)} \left( \cos\frac{\theta}{2} \right)^{n - w_H(v)} \left( \sum_{c \in \MCC} (-1)^{cv^T} Z(c) \right) \right] \dket{\theta}^{\otimes n} \\
  & = \frac{1}{|\MCC|}  \left[  \sum_{v \in \mathbb{Z}_2^n} \psi_v \left( \sin\frac{\theta}{2} \right)^{w_H(v)} \left( \cos\frac{\theta}{2} \right)^{n - w_H(v)} \left( \sum_{c \in \MCC} \sum_{z \in \mathbb{Z}_2^n} (-1)^{c (v \oplus z)^T} \dketbra{z} \right) \right] \dket{\theta}^{\otimes n} \ \ (\text{see\ Section~\ref{sec:psc}}) \\
  & = \frac{1}{|\MCC|}  \sum_{v \in \mathbb{Z}_2^n} \psi_v \left( \sin\frac{\theta}{2} \right)^{w_H(v)} \left( \cos\frac{\theta}{2} \right)^{n - w_H(v)} \left( \sum_{c \in \MCC} \sum_{z \in \mathbb{Z}_2^n} (-1)^{c (v \oplus z)^T} \left( \sin\frac{\theta}{2} \right)^{w_H(z)} \left( \cos\frac{\theta}{2} \right)^{n - w_H(z)} \dket{z} \right) \\
  & = \frac{1}{|\MCC|}  \sum_{v,z \in \mathbb{Z}_2^n} \psi_v \left( \sin\frac{\theta}{2} \right)^{w_H(v) + w_H(z)} \left( \cos\frac{\theta}{2} \right)^{2n - (w_H(v) + w_H(z))} \dket{z} \cdot \left( \sum_{c \in \MCC} (-1)^{c (v \oplus z)^T} \right) \\
  & = \frac{1}{|\MCC|}  \sum_{v \in \mathbb{Z}_2^n} \sum_{z \in \mathbb{Z}_2^n: (v \oplus z) \in \MCC^{\perp}} \psi_v \left( \sin\frac{\theta}{2} \right)^{w_H(v) + w_H(z)} \left( \cos\frac{\theta}{2} \right)^{2n - (w_H(v) + w_H(z))} \dket{z} \cdot \left( |\MCC| \right) \\
  & = \sum_{v \in \mathbb{Z}_2^n} \sum_{z \in (v \oplus \MCC^{\perp})} \psi_v \left( \sin\frac{\theta}{2} \right)^{w_H(v) + w_H(z)} \left( \cos\frac{\theta}{2} \right)^{2n - (w_H(v) + w_H(z))} \dket{z} \\
  & = \sum_{m \in \mathbb{Z}_2^k} \sum_{v \in (m G_{\MCC^{\topbot}} \oplus \MCC^{\perp})} \sum_{z \in (v \oplus \MCC^{\perp})} \psi_v \left( \sin\frac{\theta}{2} \right)^{w_H(v) + w_H(z)} \left( \cos\frac{\theta}{2} \right)^{2n - (w_H(v) + w_H(z))} \dket{z} \\
  & = \sum_{m \in \mathbb{Z}_2^k} \left( \sum_{v \in (m G_{\MCC^{\topbot}} \oplus \MCC^{\perp})} \psi_v \left( \sin\frac{\theta}{2} \right)^{w_H(v)} \left( \cos\frac{\theta}{2} \right)^{n - w_H(v)} \right) \left( \sum_{z \in ( m G_{\MCC^{\topbot}} \oplus \MCC^{\perp} )} \left( \sin\frac{\theta}{2} \right)^{w_H(z)} \left( \cos\frac{\theta}{2} \right)^{n - w_H(z)} \dket{z} \right).
\end{align}
Now consider a specific $m = \hat{m} \in \mathbb{Z}_2^k$ and set
\begin{align}
\psi_v & \coloneqq 
\begin{cases}
\dfrac{\left( \sin\frac{\theta}{2} \right)^{w_H(v)} \left( \cos\frac{\theta}{2} \right)^{n - w_H(v)}}{ \sqrt{ \sum_{y \in ( \hat{m} G_{\MCC^{\top}} \oplus \MCC )} \left( \sin^2\frac{\theta}{2} \right)^{w_H(y)} \left( \cos^2\frac{\theta}{2} \right)^{n - w_H(y)} } } & \ \text{if}\ v \in (\hat{m} G_{\MCC^{\topbot}} \oplus \MCC^{\perp}), \\
0 & \ \text{if}\ v \notin (\hat{m} G_{\MCC^{\topbot}} \oplus \MCC^{\perp})
\end{cases} \\
\label{eq:alpha_v_psi}
  & = 
\begin{cases}
\dfrac{ \sqrt{ p^{w_H(v)} \left( 1-p \right)^{n - w_H(v)} } }{ \sqrt{ \sum_{y \in ( \hat{m} G_{\MCC^{\topbot}} \oplus \MCC )} p^{w_H(y)} \left( 1-p \right)^{n - w_H(y)} } } & \ \text{if}\ v \in (\hat{m} G_{\MCC^{\topbot}} \oplus \MCC^{\perp}), \\
0 & \ \text{if}\ v \notin (\hat{m} G_{\MCC^{\topbot}} \oplus \MCC^{\perp})
\end{cases} \\
%
\Rightarrow \dket{\psi} = \sum_{v \in \mathbb{Z}_2^n} \psi_v \dket{v} & =  \sum_{v \in (\hat{m} G_{\MCC^{\topbot}} \oplus \MCC^{\perp})} \frac{ \sqrt{ p^{w_H(v)} \left( 1-p \right)^{n - w_H(v)} } }{ \sqrt{ \sum_{y \in ( \hat{m} G_{\MCC^{\topbot}} \oplus \MCC^{\perp} )} p^{w_H(y)} \left( 1-p \right)^{n - w_H(y)} } } \dket{v}.
\end{align}
Then we observe that
\begin{align}
\rho \dket{\psi} & = \left( \sum_{v \in (\hat{m} G_{\MCC^{\topbot}} \oplus \MCC^{\perp})} \left( \sin^2\frac{\theta}{2} \right)^{w_H(v)} \left( \cos^2\frac{\theta}{2} \right)^{n - w_H(v)} \right) \left( \sum_{z \in (\hat{m} G_{\MCC^{\topbot}} \oplus \MCC^{\perp})} \frac{\left( \sin\frac{\theta}{2} \right)^{w_H(z)} \left( \cos\frac{\theta}{2} \right)^{n - w_H(z)}}{ \sqrt{ \sum_{y \in ( \hat{m} G_{\MCC^{\topbot}} \oplus \MCC^{\perp} )} p^{w_H(y)} \left( 1-p \right)^{n - w_H(y)} } } \dket{z} \right) \\
%
  & = \left( \sum_{v \in (\hat{m} G_{\MCC^{\topbot}} \oplus \MCC^{\perp})} p^{w_H(v)} \left( 1 - p \right)^{n - w_H(v)} \right) \dket{\psi}.
\end{align}
Thus, we have obtained $2^k$ eigenvalues and eigenvectors for $\rho$ by considering all $\hat{m} \in \mathbb{Z}_2^k$.
Since $\rho$ is proportional to the sum of exactly $2^k$ independent rank-$1$ projectors $Z(c) \dketbra{\theta}^{\otimes n} Z(c)$, corresponding to all $c \in \MCC$, we have calculated all the non-zero eigenvalues and the corresponding eigenvectors.
\end{IEEEproof}

Therefore, we have shown that
\begin{align}
\rho & = \frac{1}{|\MCC|} \sum_{c \in \MCC} Z(c) \dketbra{\theta}^{\otimes n} Z(c) \\
  & = \sum_{m \in \mathbb{Z}_2^k} \lambda(m) \dketbra{\psi(m)} \\
  & = \left( \sum_{m \in \mathbb{Z}_2^k} \sqrt{\lambda(m)} \dketbra{\psi(m)} \right) \left( \sum_{m \in \mathbb{Z}_2^k} \sqrt{\lambda(m)} \dketbra{\psi(m)} \right)^{\dagger} \\
  & \eqqcolon \phi \phi^{\dagger},
\end{align}
so that we have determined the \emph{factor} of $\rho$ to be $\phi = \sum_{m \in \mathbb{Z}_2^k} \sqrt{\lambda(m)} \dketbra{\psi(m)}$ (as per the definition of factor in~\cite{Eldar-it04}).
As stated before, the full set of hypothesis states is given by $\{ \rho_h =  Z(y_h)\, \rho\, Z(y_h) ; \ h \in \mathbb{Z}_2^{n-k} \}$, where $y_h \coloneqq h G_{\MCC^{\top}}$.

\begin{remark}
Note that we can exactly relate these eigenvalues to the 
secrecy problem on the BSC since $\lambda(m) = 2^{-k/2} \hat{s}(m)$ based on~\eqref{eq:shat_final}.
Hence, the density matrices associated with the secrecy problem on the $\psc(\theta)$, when using the cosets of $\MCC$, are diagonal in an eigenbasis that can be bijectively related to the cosets of $\MCC^{\perp}$ in $\mathbb{Z}_2^n$, and the eigenvalues form the posterior distribution for the secrecy problem on the $\bsc\left( \frac{1-\cos\theta}{2} \right)$ while employing the cosets of $\MCC^{\perp}$. 
Furthermore, these eigenvalues also form the eigenvalues of the average output state for channel coding on the $\psc(\theta)$ (see Lemma~\ref{lem:cc_psc_Gram_eigenvalues}), and provide the eigenvalues of the average output state for channel coding on the $\bsc\left( \frac{1-\cos\theta}{2} \right)$~\eqref{eq:cc_bsc_density_matrices}.
\end{remark}

Let us calculate the relevant fidelity parameter for this secrecy problem over the PSC.
The CQ state for this state discrimination problem is given by
\begin{align}
\Psi_{\hat{A} B^n} & = \frac{1}{2^{n-k}} \sum_{h \in \mathbb{Z}_2^{n-k}} \dketbra{h}_{\hat{A}} \otimes \left[ Z(y_h)\ \rho\ Z(y_h) \right]_{B^n} \\
  & = \frac{1}{2^{n-k}} \sum_{h \in \mathbb{Z}_2^{n-k}} \dketbra{h}_{\hat{A}} \otimes \left[ \sum_{m \in \mathbb{Z}_2^k} \lambda(m) Z(y_h) \dketbra{\psi(m)} Z(y_h) \right]_{B^n} \\
  & = \sum_{h \in \mathbb{Z}_2^{n-k}} \sum_{m \in \mathbb{Z}_2^k} \frac{\lambda(m)}{2^{n-k}} \dketbra{h}_{\hat{A}} \otimes \left[ Z(y_h) \dketbra{\psi(m)} Z(y_h) \right]_{B^n}.
\end{align}
Observe that this is an eigendecomposition of the CQ state $\Psi_{\hat{A} B^n}$.
This state can be related to the CQ state in~\eqref{eq:coding_bsc_cq_state} for channel coding on the BSC by duality (see~\cite{Renes-it18}).
The entropy we are interested in is given by
\begin{align}
\label{eq:secrecy_psc_fidelity}
H_{\max}(\hat{Z}|B^n)_{\Psi} & \coloneqq \log|\hat{A}| + \underset{\sigma \in \mathcal{D}(\mathcal{H}_{B^n})}{\max} \ \log \mathcal{F}\left( \Psi_{\hat{A} B^n}, \frac{1}{2^{n-k}} \mathbb{I}_{\hat{A}} \otimes \sigma_{B^n} \right),
\end{align}
where the fidelity is defined as $\mathcal{F}(\rho,\sigma) \coloneqq \norm{\sqrt{\rho} \sqrt{\sigma}}_1^2$, and $\norm{A}_1 \coloneqq \tr{\sqrt{A^{\dagger} A}}$ is the trace norm of the matrix $A$.
The maximization is taken over all density matrices in the subsystem $B^n$, i.e., in the space $\mathbb{C}^{2^n \times 2^n}$.

\begin{theorem}
\label{thm:psc_secrecy_fidelity}
Define $\beta(v) \coloneqq 2^{n-k} p_{v^*}/q = 2^{n-k} p^{w_H(v^*)} (1 - p)^{n - w_H(v^*)}/q$, where $q \coloneqq \sum_{v \in \MCC^{\topbot}} p_{v^*}$ (see Def.~\ref{def:min_wt_vec}).
Then, the optimal choice for $\sigma_{B^n}$ that equates the fidelity in~\eqref{eq:secrecy_psc_fidelity} to the MAP probability of success on BSC is 
\begin{align}
\sigma_{B^n} = \frac{1}{2^{n-k}} \tilde{\sigma}_{B^n}, \ \ 
\tilde{\sigma}_{B^n} & \coloneqq \sum_{v \in \MCC^{\topbot}} \beta(v) \dketbra{v^*}.
\end{align}
\end{theorem}
\begin{IEEEproof}
Recollecting that $\lambda(m) = \sum_{u \in m G_{\MCC^{\topbot}} \oplus \MCC^{\perp}} p_u$, we write $\dket{\psi(m)} = \sum_{u \in m G_{\MCC^{\topbot}} \oplus \MCC^{\perp}} \sqrt{\dfrac{p_u}{\lambda(m)}} \dket{u}$.
Defining the vectors $v(m) \coloneqq m G_{\MCC^{\topbot}}$, we calculate
\begin{IEEEeqnarray}{rCl+x*}
&  & \mathcal{F}\left( \Psi_{\hat{A} B^n}, \frac{1}{2^{n-k}} \mathbb{I}_{\hat{A}} \otimes \frac{1}{2^{n-k}} \tilde{\sigma}_{B^n} \right) \nonumber \\
  & = & \norm{\sqrt{\Psi_{\hat{A} B^n}} \sqrt{\frac{1}{2^{n-k}} \mathbb{I}_{\hat{A}} \otimes \frac{1}{2^{n-k}} \tilde{\sigma}_{B^n} }}_1^2 \\
  & = & \norm{ \left[ \sum_{h \in \mathbb{Z}_2^{n-k}} \sum_{m \in \mathbb{Z}_2^k} \sqrt{\frac{\lambda(m)}{2^{n-k}}} \dketbra{h}_{\hat{A}} \otimes \left[ Z(y_h) \dketbra{\psi(m)} Z(y_h) \right]_{B^n} \right] \cdot \left[ \frac{1}{2^{n-k}} \mathbb{I}_{\hat{A}} \otimes \sum_{v \in \MCC^{\topbot}} \sqrt{\beta(v)} \dketbra{v^*}_{B^n} \right] }_1^2 \\
  & = & \norm{ \frac{1}{2^{n-k}} \sum_{h \in \mathbb{Z}_2^{n-k}} \sum_{m \in \mathbb{Z}_2^k} (-1)^{v(m)^* y_h^T} \sqrt{\frac{\lambda(m) 2^{n-k} p_{v(m)^*}}{2^{n-k} q}} \sqrt{\frac{p_{v(m)^*}}{\lambda(m)}} \dketbra{h}_{\hat{A}} \otimes Z(y_h) \dketbra{\psi(m)}{v(m)^*}_{B^n} }_1^2 \\
  & = & \norm{ \frac{1}{2^{n-k} \sqrt{q}} \sum_{h \in \mathbb{Z}_2^{n-k}} \sum_{m \in \mathbb{Z}_2^k} p_{v(m)^*} \dketbra{h}_{\hat{A}} \otimes Z(y_h) \dketbra{\psi(m)}{v(m)^*}_{B^n} }_1^2 \ \ (\text{since}\ y_h \in \MCC^{\top}, v(m)^* \in \MCC^{\topbot}) \\
  & = & \left( \text{Tr}\left[ \frac{1}{2^{n-k} \sqrt{q}} \left( \sum_{h \in \mathbb{Z}_2^{n-k}} \sum_{m \in \mathbb{Z}_2^k}  p_{v(m)^*}^2 \dketbra{h}_{\hat{A}} \otimes \dketbra{v(m)^*}_{B^n} \right)^{1/2} \right] \right)^2 \\
  & = & \left( \text{Tr}\left[ \frac{1}{2^{n-k} \sqrt{q}} \sum_{h \in \mathbb{Z}_2^{n-k}} \sum_{m \in \mathbb{Z}_2^k}  p_{v(m)^*} \dketbra{h}_{\hat{A}} \otimes \dketbra{v(m)^*}_{B^n} \right] \right)^2 \\
  & = & \left( \frac{1}{2^{n-k} \sqrt{q}} \sum_{h \in \mathbb{Z}_2^{n-k}} \sum_{m \in \mathbb{Z}_2^k}  p_{v(m)^*} \right)^2 \\
  & = & \left( \frac{1}{\sqrt{q}} \sum_{m \in \mathbb{Z}_2^k}  p_{v(m)^*} \right)^2 \\
  & = & q \\
  & = & \sum_{m \in \mathbb{Z}_2^k} \underset{u \in m G_{\MCC^{\topbot}} \oplus \MCC^{\perp}}{\max}\, p^{w_H(u)} (1 - p)^{n - w_H(u)} \\
  & = & \mathbb{P}\left[ \text{MAP success for $\MCC^{\perp}$ on BSC$(p)$} \right]. & \nonumber \IEEEQEDhere
\end{IEEEeqnarray}
\end{IEEEproof}

As discussed at the beginning of Section~\ref{sec:duality_coding_secrecy}, the entropic duality satisfied by a CQ channel $W$ and its dual CQ channel $W^{\perp}$ is $H_{\min}(W) + H_{\max}(W^{\perp}) = \log |\mathcal{X}|$, where $|\mathcal{X}|$ is the size of the input alphabet for both channels.
Here, we have $W = \bsc(p)$ and $W^{\perp} = \psc(\theta)$ with $p = (1 - \cos\theta)/2$.
If one performed standard channel coding over $W$, then the optimal block success probability is given by $P(W) = 2^{-H_{\min}(W)} = \mathbb{P}\left[ \text{MAP\ Success\ Prob.} \right]$.
Similarly, if one performed secret communications over $W^\perp$, then $Q(W^\perp) = \frac{1}{|\mathcal{X}|} 2^{H_{\max}(W^\perp)}$ measures the optimal decoupling between the intercepted information and the actual secret message.
Since $|\mathcal{X}| = |\hat{A}| = 2^{n-k}$ in~\eqref{eq:secrecy_psc_fidelity}, where $H_{\max}(W^\perp) = H_{\max}(\hat{Z}|B^n)_{\Psi}$, the above theorem exactly verifies the above entropic duality, or equivalently $P(W) = Q(W^{\perp})$, for this setting.

Given the expression in~\eqref{eq:secrecy_psc_fidelity}, a natural guess for the maximizer is $\sigma_{B^n} = \text{Tr}_{\hat{A}}\left[ \Psi_{\hat{A} B^n} \right] = \frac{1}{2^{n-k}} \sum_{h \in \mathbb{Z}_2^{n-k}} Z(y_h)\, \rho\, Z(y_h)$.
However, this turns out to be a suboptimal choice, and only yields a lower bound on the optimal fidelity derived above.

\begin{theorem}
\label{thm:secrecy_psc_fidelity_lbound}
The fidelity achieved by the choice $\sigma_{B^n} = \text{Tr}_{\hat{A}}\left[ \Psi_{\hat{A} B^n} \right]$ is
\begin{align}
\mathcal{F}\left( \Psi_{\hat{A} B^n}, \frac{1}{2^{n-k}} \mathbb{I}_{\hat{A}} \otimes \frac{1}{2^{n-k}} \left( \Phi \Phi^{\dagger} \right)_{B^n} \right) = \left( \sum_{m \in \mathbb{Z}_2^k} \sqrt{ \sum_{u \in (m G_{\MCC^{\topbot}} \oplus \MCC^{\perp})} \left( p^{w_H(u)} \left( 1-p \right)^{n - w_H(u)} \right)^2 } \right)^2.
\end{align}
\end{theorem}
\begin{IEEEproof}
See Appendix~\ref{sec:proof_secrecy_psc_fidelity_lbound}.
\end{IEEEproof}

A more trivial lower bound for the fidelity based on $\sigma_{B^n} = \frac{1}{2^{n}} \mathbb{I}_{2^n}$ is
\begin{align}
\mathcal{F}\left( \Psi_{\hat{A} B^n}, \frac{1}{2^{n-k}} \mathbb{I}_{\hat{A}} \otimes \frac{1}{2^{n}} \mathbb{I}_{B^n} \right)^2 & = \left( \frac{1}{\sqrt{2^k}} \sum_{m \in \mathbb{Z}_2^k} \sqrt{\frac{ \sum_{u \in (m G_{\MCC^{\topbot}} \oplus \MCC^{\perp})} p^{w_H(u)} \left( 1-p \right)^{n - w_H(u)} }{2^{n-k}}} \right)^2 \\
  & = \frac{1}{2^n} \left( \sum_{m \in \mathbb{Z}_2^k} \sqrt{ \sum_{u \in (m G_{\MCC^{\topbot}} \oplus \MCC^{\perp})} p^{w_H(u)} \left( 1-p \right)^{n - w_H(u)} } \right)^2.
\end{align}
However, even these two lower bounds still appear to require full coset information, and hence are not easier to compute than the optimal fidelity. \\

\noindent \textbf{Square Root Measurement (SRM) Does Not Maximize the Fidelity} \\

First, let us recollect the precise connection between the quantum fidelity and the classical Bhattacharyya distance~\cite[Chapter 9]{Wilde-2013}.
Let $\rho$ and $\sigma$ be two candidate states and let $\mathcal{F}(\rho,\sigma)$ be their fidelity.
For any POVM $\{ \Pi_i\, ;\ i = 1,2,\ldots,t \}$ that is used to distinguish them, there are two induced classical probability distributions: \{ $p_i = \tr{\rho \cdot \Pi_i}$ \} for $\rho$ and \{ $q_i = \tr{\sigma \cdot \Pi_i}$ \} for $\sigma$.
Since a POVM is the most general apparatus for distinguishing the states, the task of distinguishing the quantum states is equivalent to  the task of distinguishing the distributions $\{ p_i \}$ and $\{ q_i \}$, which is characterized by
\begin{align}
\mathcal{F}(\{ p_i \}, \{ q_i \}) \coloneqq \mathcal{B}(\{ p_i \}, \{ q_i \})^2 = \left( \sum_{i=1}^t \sqrt{p_i q_i} \right)^2.
\end{align}
Therefore, the fidelity between the quantum states can be expressed as
\begin{align}
\mathcal{F}(\rho, \sigma) = \underset{\{ \Pi_i \}}{\min}\ \mathcal{F}(\{ p_i \}, \{ q_i \}),
\end{align}
where the minimization is over all possible POVMs $\{ \Pi_i \}$.

Since we have obtained the factors $\phi_h = Z(y_h) \phi$ for the hypothesis states, we can determine the fidelity induced by the SRM while acting on the candidate states in the optimal fidelity expression above (Theorem~\ref{thm:psc_secrecy_fidelity}).
Given the uniform prior assumption on the secret message $h$, the SRM in this scenario is given by the POVM $\{ \Pi_h = \mu_h \mu_h^{\dagger},\ h \in \mathbb{Z}_2^{n-k} \}$, where 
\begin{align}
\mu_h & \coloneqq \left( \Phi \Phi^{\dagger} \right)^{-1/2} \phi_h, \\
\Phi & \coloneqq 
\left[
\begin{array}{c|c|c|c|c|c}
\phi_{00\cdots 0} & \phi_{00\cdots 1} & \cdots & \phi_h & \cdots & \phi_{11\cdots 1}
\end{array} 
\right].
\end{align}

\begin{lemma}
The matrix $\Phi \Phi^{\dagger}$ is diagonal in the computational basis and can be written as
\begin{align}
\Phi \Phi^{\dagger} = \sum_{v \in \mathbb{Z}_2^n} \alpha(v) \dketbra{v},
\end{align}
where $\alpha(v) = 2^{n-k} p_v = 2^{n-k} p^{w_H(v)} \left( 1-p \right)^{n - w_H(v)}$.
\end{lemma}
\begin{IEEEproof}
We can calculate the action of $\Phi \Phi^{\dagger}$ on an arbitrary state $\dket{\gamma} = \sum_{v \in \mathbb{Z}_2^n} \gamma_v \dket{v}$ as follows.
\begin{align}
\Phi \Phi^{\dagger} \cdot \dket{\gamma} & = \sum_{h \in \mathbb{Z}_2^{n-k}} \phi_h \phi_h^{\dagger} \cdot \sum_{v \in \mathbb{Z}_2^n} \gamma_v \dket{v} \\
  & = \sum_{h \in \mathbb{Z}_2^{n-k}} \sum_{m \in \mathbb{Z}_2^k} \lambda(m) Z(y_h) \dketbra{\psi(m)} Z(y_h) \cdot \sum_{v \in \mathbb{Z}_2^n} \gamma_v \dket{v} \\
  & = \sum_{y_h \in \MCC^{\top}} \sum_{m \in \mathbb{Z}_2^k} \sum_{v \in \mathbb{Z}_2^n} \gamma_v \lambda(m) Z(y_h) \dket{\psi(m)} (-1)^{y_h v^T} \dbraket{\psi(m)}{v} \\
  & = \sum_{y_h \in \MCC^{\top}} \sum_{m \in \mathbb{Z}_2^k} \sum_{v \in (m G_{\MCC^{\topbot}} \oplus \MCC^{\perp})} \gamma_v  (-1)^{y_h v^T}  \lambda(m)   \frac{\left( \sin\frac{\theta}{2} \right)^{w_H(v)} \left( \cos\frac{\theta}{2} \right)^{n - w_H(v)}}{\sqrt{\lambda(m)}}  Z(y_h) \dket{\psi(m)} \\
  & = \sum_{y_h \in \MCC^{\top}} \sum_{m \in \mathbb{Z}_2^k} \sum_{v \in (m G_{\MCC^{\topbot}} \oplus \MCC^{\perp})} \gamma_v  (-1)^{y_h v^T}  \lambda(m)   \frac{\left( \sin\frac{\theta}{2} \right)^{w_H(v)} \left( \cos\frac{\theta}{2} \right)^{n - w_H(v)}}{\sqrt{\lambda(m)}} \nonumber \\
  & \hspace{4.5cm} \sum_{z \in (m G_{\MCC^{\topbot}} \oplus \MCC^{\perp})} (-1)^{y_h z^T} \frac{ \left( \sin\frac{\theta}{2} \right)^{w_H(z)} \left( \cos\frac{\theta}{2} \right)^{n - w_H(z)}}{\sqrt{\lambda(m)}} \dket{z} \\
  & = \sum_{y_h \in \MCC^{\top}} \sum_{m \in \mathbb{Z}_2^k} \sum_{v,z \in (m G_{\MCC^{\topbot}} \oplus \MCC^{\perp})} \gamma_v (-1)^{y_h (v \oplus z)^T} \left( \sin\frac{\theta}{2} \right)^{w_H(v) + w_H(z)} \left( \cos\frac{\theta}{2} \right)^{2n - w_H(v) - w_H(z)} \dket{z} \\
  & =  \sum_{m \in \mathbb{Z}_2^k} \sum_{v,z \in (m G_{\MCC^{\topbot}} \oplus \MCC^{\perp})} \gamma_v \left( \sin\frac{\theta}{2} \right)^{w_H(v) + w_H(z)} \left( \cos\frac{\theta}{2} \right)^{2n - w_H(v) - w_H(z)} \dket{z} \left( \sum_{y_h \in \MCC^{\top}} (-1)^{y_h (v \oplus z)^T} \right) \\
  & \overset{\text{(a)}}{=} \sum_{m \in \mathbb{Z}_2^k} \sum_{v \in (m G_{\MCC^{\topbot}} \oplus \MCC^{\perp})} \gamma_v \left( \sin^2\frac{\theta}{2} \right)^{w_H(v)} \left( \cos^2\frac{\theta}{2} \right)^{n - w_H(v)} \dket{v} \cdot \left( |\MCC^{\top}| \right) \\
  & = \left|\MCC^{\top}\right| \sum_{m \in \mathbb{Z}_2^k} \sum_{v \in (m G_{\MCC^{\topbot}} \oplus \MCC^{\perp})} \gamma_v\  p^{w_H(v)} \left( 1-p \right)^{n - w_H(v)} \dket{v} \\
  & = 2^{n-k} \sum_{v \in \mathbb{Z}_2^n} \gamma_v\  p^{w_H(v)} \left( 1-p \right)^{n - w_H(v)} \dket{v}.
\end{align}
In step (a), note that $(v \oplus z) \in \MCC^{\perp}$ but for the last inner summation to be non-zero we also need $(v \oplus z) \in \MCC^{\topbot}$, which implies that we need $v \oplus z = 0$.
Hence, we see that $\Phi \Phi^{\dagger}$ is diagonal in the standard basis $\{ \dket{v},\ v \in \mathbb{Z}_2^n \}$ with eigenvalues $\{ \alpha(v) = 2^{n-k} p^{w_H(v)} \left( 1-p \right)^{n - w_H(v)} \}$.
It can also be seen from the original expression of $\Phi \Phi^{\dagger}$ that 
\begin{align}
\text{Tr}\left( \Phi \Phi^{\dagger} \right) & = \sum_{h \in \mathbb{Z}_2^{n-k}} \tr{ \phi_h \phi_h^{\dagger} } = \sum_{h \in \mathbb{Z}_2^{n-k}} \tr{ \rho_h } = \sum_{h \in \mathbb{Z}_2^{n-k}} (1) = 2^{n-k} = \sum_{v \in \mathbb{Z}_2^n} \alpha(v).
\end{align}
Hence, this verifies that the eigenvalues $\alpha(v)$ produce the correct trace.
\end{IEEEproof}

Therefore, in order to obtain the SRM measurement operators, we can further calculate
\begin{align}
\left( \Phi \Phi^{\dagger} \right)^{-1/2} & = \left( \sum_{h \in \mathbb{Z}_2^{n-k}} \sum_{m \in \mathbb{Z}_2^k} \lambda(m) Z(y_h) \dketbra{\psi(m)} Z(y_h) \right)^{-1/2} \\
  & = \left( \sum_{v \in \mathbb{Z}_2^n} \alpha(v) \dketbra{v} \right)^{-1/2} \\
  & = \sum_{v \in \mathbb{Z}_2^n} \frac{1}{\sqrt{\alpha(v)}} \dketbra{v}.
\end{align}
Now, we observe that
\begin{align}
\phi^{\dagger} \left( \Phi \Phi^{\dagger} \right)^{-1/2} \phi & =  \sum_{m \in \mathbb{Z}_2^k} \sqrt{\lambda(m)} \dketbra{\psi(m)} \cdot \sum_{v \in \mathbb{Z}_2^n} \frac{1}{\sqrt{\alpha(v)}} \dketbra{v} \cdot \sum_{m' \in \mathbb{Z}_2^k} \sqrt{\lambda(m')} \dketbra{\psi(m')} \\
  & = \sum_{m \in \mathbb{Z}_2^k} \sum_{v \in (m G_{\MCC^{\topbot}} \oplus \MCC^{\perp})} \frac{\lambda(m)}{\sqrt{\alpha(v)}} \  |\dbraket{\psi(m)}{v}|^2 \ \dketbra{\psi(m)} \\
  & = \frac{1}{\sqrt{2^{n-k}}} \sum_{m \in \mathbb{Z}_2^k} \left( \sum_{v \in (m G_{\MCC^{\topbot}} \oplus \MCC^{\perp})} \sqrt{p^{w_H(v)} \left( 1-p \right)^{n - w_H(v)}} \right) \dketbra{\psi(m)}. 
\end{align}
In general, this is again not a scalar multiple of the identity, just like we observed earlier for the channel coding over BSC problem.
Hence, we are unable to use~\cite[Theorem 3]{Eldar-it04} to conclude whether the square root measurement maximizes the probability of success in this hypothesis testing problem of distinguishing $\{ \rho_h,\ h \in \mathbb{Z}_2^{n-k} \}$. 

Nevertheless, we are only interested in the Bhattacharyya distance induced by the SRM when acting on the two density matrices in the optimal fidelity expression from Theorem~\ref{thm:psc_secrecy_fidelity},
\begin{align}
\mathcal{F}\left( \Psi_{\hat{A} B^n}, \frac{1}{2^{n-k}} \mathbb{I}_{\hat{A}} \otimes \frac{1}{2^{n-k}} \tilde{\sigma}_{B^n} \right),\ \ \tilde{\sigma}_{B^n} \coloneqq \sum_{v \in \MCC^{\topbot}} \beta(v) \dketbra{v^*}.
\end{align}

\begin{theorem}
\label{thm:psc_secrecy_srm_suboptimal}
Let $d_{h'} = \tr{ \Psi_{\hat{A} B^n} \cdot \left( \mathbb{I}_{\hat{A}} \otimes \Pi_{h'} \right) }$ and $f_{h'} = \tr{ \left( \frac{1}{2^{n-k}} \mathbb{I}_{\hat{A}} \otimes \frac{1}{2^{n-k}} \tilde{\sigma}_{B^n} \right) \cdot \left( \mathbb{I}_{\hat{A}} \otimes \Pi_{h'} \right) }$ be the classical distributions induced by the SRM.
Then, each of these is the uniform distribution and hence, the SRM is highly suboptimal for secrecy over the PSC. 
\end{theorem}
\begin{IEEEproof}
Recall that the POVM for the SRM in this case was $\{ \Pi_h = \mu_h \mu_h^{\dagger},\ h \in \mathbb{Z}_2^{n-k} \}$, where 
\begin{align}
\mu_h & \coloneqq \left( \Phi \Phi^{\dagger} \right)^{-1/2} \phi_h \\
  & = \sum_{v \in \mathbb{Z}_2^n} \frac{1}{\sqrt{\alpha(v)}} \dketbra{v} \cdot Z(y_h) \sum_{m \in \mathbb{Z}_2^k} \sqrt{\lambda(m)} \dketbra{\psi(m)} \\
  & = Z(y_h) \cdot \sum_{v \in \mathbb{Z}_2^n} \frac{1}{\sqrt{\alpha(v)}} \dketbra{v} \cdot \phi.
%
%
\end{align}
Hence, the SRM operators are also geometrically uniform (GU), i.e., $\Pi_h = Z(y_h) \Pi_0 Z(y_h)$, just like the candidate states, i.e., $\rho_h = Z(y_h) \rho Z(y_h)$.
First, the probability distribution induced via $\Psi_{\hat{A} B^n}$ is
\begin{align}
d_{h'} & = \tr{ \Psi_{\hat{A} B^n} \cdot \left( \mathbb{I}_{\hat{A}} \otimes \Pi_{h'} \right) } \\
  & = \frac{1}{2^{n-k}} \sum_{h \in \mathbb{Z}_2^{n-k}} \tr{ \dketbra{h} } \otimes \tr{ \rho_h \cdot \Pi_{h'} } \\
  & = \frac{1}{2^{n-k}} \sum_{h \in \mathbb{Z}_2^{n-k}} \tr{ \rho \cdot \Pi_{h \oplus h'} } \ \ (\text{GU\ symmetry}) \\
  & = \frac{1}{2^{n-k}},
\end{align}
which is the uniform distribution over all messages.
Next, the probability distribution induced via $\frac{1}{2^{n-k}} \mathbb{I}_{\hat{A}} \otimes \frac{1}{2^{n-k}} \tilde{\sigma}_{B^n}$ is
\begin{align}
f_{h'} & = \tr{ \left( \frac{1}{2^{n-k}} \mathbb{I}_{\hat{A}} \otimes \frac{1}{2^{n-k}} \tilde{\sigma}_{B^n} \right) \cdot \left( \mathbb{I}_{\hat{A}} \otimes \Pi_{h'} \right) } \\
  & = \tr{ \frac{1}{2^{n-k}} \mathbb{I}_{\hat{A}} } \cdot \frac{1}{2^{n-k}} \tr{ \sum_{v \in \MCC^{\topbot}} \beta(v) \dketbra{v^*} \cdot \sum_{v \in \mathbb{Z}_2^n} \frac{1}{\sqrt{\alpha(v)}} \dketbra{v} \cdot \rho_{h'} \cdot \sum_{v \in \mathbb{Z}_2^n} \frac{1}{\sqrt{\alpha(v)}} \dketbra{v} } \\
  & = \frac{1}{2^{n-k}} \tr{ \left( \sum_{v \in \mathbb{Z}_2^n} \frac{1}{\sqrt{\alpha(v)}} \dketbra{v} \cdot \sum_{v \in \MCC^{\topbot}} \beta(v) \dketbra{v^*} \cdot \sum_{v \in \mathbb{Z}_2^n} \frac{1}{\sqrt{\alpha(v)}} \dketbra{v} \right) \cdot \rho_{h'} } \\
  & = \frac{1}{2^{n-k}} \tr{ \sum_{v \in \MCC^{\topbot}} \frac{\beta(v)}{\alpha(v^*)} \dketbra{v^*} \cdot \rho_{h'} } \\
   & = \frac{1}{2^{n-k}} \tr{ \sum_{v \in \MCC^{\topbot}} \frac{2^{n-k} p_{v^*}/q}{2^{n-k} p_{v^*}} \dketbra{v^*} \cdot Z(y_{h'}) \sum_{m \in \mathbb{Z}_2^k} \lambda(m) \dketbra{\psi(m)} Z(y_{h'}) } \\
  & = \frac{1}{2^{n-k} q} \tr{ Z(y_{h'}) \sum_{v \in \MCC^{\topbot}} \dketbra{v^*} \cdot Z(y_{h'}) \sum_{m \in \mathbb{Z}_2^k} \lambda(m) \dketbra{\psi(m)} } \\
  & = \frac{1}{2^{n-k} q} \sum_{v \in \MCC^{\topbot}} \sum_{m \in \mathbb{Z}_2^k} \lambda(m) \left| \dbraket{v^*}{\psi(m)} \right|^2\ \ ; \ \ \dket{\psi(m)} = \sum_{u \in m G_{\MCC^{\topbot}} \oplus \MCC^{\perp}} \sqrt{\dfrac{p_u}{\lambda(m)}} \dket{u} \\
  & = \frac{1}{2^{n-k} q} \sum_{v \in \MCC^{\topbot}, v = \hat{m} G_{\MCC^{\topbot}}} \lambda(\hat{m}) \frac{p_{v^*}}{\lambda(\hat{m})} \\
  & = \frac{1}{2^{n-k} q} \sum_{v \in \MCC^{\topbot}} p_{v^*} \\
  & = \frac{1}{2^{n-k}},
\end{align}
which is also the uniform distribution.
Hence, the Bhattacharyya distance induced by the SRM is simply $\mathcal{F}(\{ d_{h'} \}, \{ f_{h'} \}) = 1$.
\end{IEEEproof}

Therefore, the SRM is far from inducing the optimal fidelity $\mathcal{F}\left( \Psi_{\hat{A} B^n}, \frac{1}{2^{n-k}} \mathbb{I}_{\hat{A}} \otimes \frac{1}{2^{n-k}} \tilde{\sigma}_{B^n} \right)$ when acting on the constituent states in the expression, and, in fact, it only provides a trivial upper bound on this fidelity.
From~\cite[Theorem 3]{Eldar-it04}, we know that, if probability of success is the objective, then the optimal POVM for distinguishing $\{ \rho_h = Z(y_h)\, \rho\, Z(y_h),\ h \in \mathbb{Z}_2^{n-k} \}$ is always GU with respect to the same group $\{ Z(y_h),\ y_h \in \MCC^{\top} \}$ that generates the candidate states.
If this also extends to the goal of achieving the aforesaid optimal fidelity, then we see that the distribution $d_{h'}$ in Theorem~\ref{thm:psc_secrecy_srm_suboptimal} is always uniform.
Therefore, we are left with finding the POVM that induces the appropriate distribution $f_{h'}$, such that the resulting squared Bhattacharyya distance equals the optimal fidelity in Theorem~\ref{thm:psc_secrecy_fidelity}.

\section{Conclusion}
\label{sec:conclusion}

Renes showed that a CQ channel $W$ and its dual $W^\perp$, both with $d$ input symbols, satisfy $H(W)+H^\perp (W^\perp) = \log d$ for a pair of primal and dual entropies $H$ and $H^\perp$.
For the case where $W = \psc(\theta)$ is the CQ binary PSC with parameter $\theta$, one finds that $W^\perp = \bsc(p)$ is the classical BSC with parameter $p \coloneqq (1 - \cos\theta)/2$.
From this, one gets a duality connection between coding for error correction and wire-tap secrecy one the PSC and BSC channels.
In this paper, we provide an alternate derivation for this particular duality relationship by directly calculating closed-form expressions for both the block error rate for error correction and the Bhattacharyya distance performance metric for the secrecy problem.
Our approach also establishes some results from~\cite{Renes-it18} for Von Neumann entropy including the coding-secrecy result and GEXIT function duality.
We believe that our approach highlights some connections with known classical results and may be more accessible for researchers with a limited background in quantum information theory.
Our calculations also identify when the SRM is optimal (or suboptimal) for the considered problems.
As mentioned above, it remains open to find the optimal measurement for secrecy on the PSC.

\section*{Acknowledgment}

This work was supported in part by the National Science Foundation (NSF) under Grant No. 1908730, 1910571, and 1855879. Any opinions, findings, conclusions, and recommendations expressed in this material are those of the authors and do not necessarily reflect the views of these sponsors.



\appendices



\section{Proof of Theorem~\ref{thm:secrecy_psc_fidelity_lbound}}
\label{sec:proof_secrecy_psc_fidelity_lbound}

We calculate
\begin{align}
& \mathcal{F}\left( \Psi_{\hat{A} B^n}, \frac{1}{2^{n-k}} \mathbb{I}_{\hat{A}} \otimes \frac{1}{2^{n-k}} \left( \Phi \Phi^{\dagger} \right)_{B^n} \right)^2 \nonumber \\
  & = \norm{\sqrt{\Psi_{\hat{A} B^n}} \sqrt{\frac{1}{2^{n-k}} \mathbb{I}_{\hat{A}} \otimes \frac{1}{2^{n-k}} \left( \Phi \Phi^{\dagger} \right)_{B^n} }}_1^2 \\
  & = \norm{ \left[ \sum_{h \in \mathbb{Z}_2^{n-k}} \sum_{m \in \mathbb{Z}_2^k} \sqrt{\frac{\lambda(m)}{2^{n-k}}} \dketbra{h}_{\hat{A}} \otimes \left[ Z(y_h) \dketbra{\psi(m)} Z(y_h) \right]_{B^n} \right] \cdot \left[ \frac{1}{2^{n-k}} \mathbb{I}_{\hat{A}} \otimes \sum_{v \in \mathbb{Z}_2^n} \sqrt{\alpha(v)} \dketbra{v}_{B^n} \right] }_1^2 \\
  & = \norm{ \frac{1}{2^{n-k}} \sum_{h \in \mathbb{Z}_2^{n-k}} \sum_{m \in \mathbb{Z}_2^k} \sum_{v \in (m G_{\MCC^{\topbot}} \oplus \MCC^{\perp})} (-1)^{y_h v^T} \sqrt{\frac{\lambda(m) \alpha(v)}{2^{n-k}}} \frac{\left( \sin\frac{\theta}{2} \right)^{w_H(v)} \left( \cos\frac{\theta}{2} \right)^{n - w_H(v)}}{\sqrt{\lambda(m)}} \dketbra{h}_{\hat{A}} \otimes Z(y_h) \dketbra{\psi(m)}{v}_{B^n} }_1^2 \\
  & = \norm{ \frac{1}{2^{n-k}} \sum_{h \in \mathbb{Z}_2^{n-k}} \sum_{m \in \mathbb{Z}_2^k} \sum_{v \in (m G_{\MCC^{\topbot}} \oplus \MCC^{\perp})} (-1)^{y_h v^T} p^{w_H(v)} \left( 1-p \right)^{n - w_H(v)} \dketbra{h}_{\hat{A}} \otimes Z(y_h) \dketbra{\psi(m)}{v}_{B^n} }_1^2 \\
  & = \left( \text{Tr}\left[ \frac{1}{2^{n-k}} \left( \sum_{h \in \mathbb{Z}_2^{n-k}} \sum_{m \in \mathbb{Z}_2^k} \sum_{v',v \in (m G_{\MCC^{\topbot}} \oplus \MCC^{\perp})} (-1)^{y_h (v \oplus v')^T} p^{w_H(v) + w_H(v')} \left( 1-p \right)^{2n - w_H(v) - w_H(v')} \dketbra{h}_{\hat{A}} \otimes \dketbra{v'}{v}_{B^n} \right)^{1/2} \right] \right)^2 \\
  & = \left( \text{Tr}\left[ \frac{1}{2^{n-k}} \left( \sum_{h \in \mathbb{Z}_2^{n-k}} \sum_{m \in \mathbb{Z}_2^k} \left( \sum_{u \in (m G_{\MCC^{\topbot}} \oplus \MCC^{\perp})} \left( p^{w_H(u)} \left( 1-p \right)^{n - w_H(u)} \right)^2 \right) \dketbra{h}_{\hat{A}} \otimes \dketbra{\sigma(h,m)}_{B^n} \right)^{1/2} \right] \right)^2,
\end{align}
where we have defined the states
\begin{align}
\dket{\sigma(h,m)} \coloneqq \sum_{v \in (m G_{\MCC^{\topbot}} \oplus \MCC^{\perp})} \frac{ (-1)^{y_h v^T} p^{w_H(v)} \left( 1-p \right)^{n - w_H(v)} }{ \sqrt{ \sum_{u \in (m G_{\MCC^{\topbot}} \oplus \MCC^{\perp})} \left( p^{w_H(u)} \left( 1-p \right)^{n - w_H(u)} \right)^2 } } \dket{v}.
\end{align}
Note that $\{ \dket{h}_{\hat{A}} \otimes \dket{\sigma(h,m)}_{B^n} \ ;\ h \in \mathbb{Z}_2^{n-k}, m \in \mathbb{Z}_2^k \}$ form an orthonormal basis.
Hence, we complete the fidelity calculation to get
\begin{align}
& \mathcal{F}\left( \Psi_{\hat{A} B^n}, \frac{1}{2^{n-k}} \mathbb{I}_{\hat{A}} \otimes \frac{1}{2^{n-k}} \left( \Phi \Phi^{\dagger} \right)_{B^n} \right)^2 \nonumber \\
  & = \left( \text{Tr}\left[ \frac{1}{2^{n-k}} \sum_{h \in \mathbb{Z}_2^{n-k}} \sum_{m \in \mathbb{Z}_2^k} \sqrt{ \sum_{u \in (m G_{\MCC^{\topbot}} \oplus \MCC^{\perp})} \left( p^{w_H(u)} \left( 1-p \right)^{n - w_H(u)} \right)^2 } \dketbra{h}_{\hat{A}} \otimes \dketbra{\sigma(h,m)}_{B^n} \right] \right)^2 \\
  & = \left( \frac{1}{2^{n-k}} \sum_{h \in \mathbb{Z}_2^{n-k}} \sum_{m \in \mathbb{Z}_2^k} \sqrt{ \sum_{u \in (m G_{\MCC^{\topbot}} \oplus \MCC^{\perp})} \left( p^{w_H(u)} \left( 1-p \right)^{n - w_H(u)} \right)^2 } \right)^2 \\
  & = \left( \sum_{m \in \mathbb{Z}_2^k} \sqrt{ \sum_{u \in (m G_{\MCC^{\topbot}} \oplus \MCC^{\perp})} \left( p^{w_H(u)} \left( 1-p \right)^{n - w_H(u)} \right)^2 } \right)^2.
\end{align}
This completes the proof.  \hfill \IEEEQEDhere

\section{Fourier (or Factor Graph) Duality of Linear Codes}
\label{sec:fg_duality}

Let $\mathcal{X}=\left\{ 0,1,\ldots,q-1\right\} $ be a finite alphabet and consider the function $f:\mathcal{X}^{n}\to\mathbb{R}$. Now, suppose that we want to compute the sum
\begin{align}
Z(\vecnot{\mu})\triangleq\sum_{x_{1}^{n}\in\mathcal{X}^{n}}f(x_{1},x_{2},\ldots,x_{n})\prod_{j=1}^{n}\mu_{j}(x_{j})
\end{align}
for the vector $\vecnot{\mu}=(\mu_{1},\ldots,\mu_{n})$ where each element is a function $\mu_{j}:\mathcal{X}\to\mathbb{R}$. Using an invertible $q\times q$ matrix $A$, the same quantity can be written as 
\begin{align}
Z(\vecnot{\mu})=\sum_{x_{1}^{n}\in\mathcal{X}^{n}}f(x_{1},x_{2},\ldots,x_{n})\prod_{j=1}^{n}\sum_{\hat{x}\in\mathcal{X}}A_{x_{j},\hat{x}}\sum_{x\in\mathcal{X}}A_{\hat{x},x}^{-1}\mu_{j}(x).
\end{align}
If we define $\vecnot{\hat{\mu}}=(\hat{\mu}_{1},\ldots,\hat{\mu}_{n})$ by $\hat{\mu}_{j}(\hat{x}_{j})\triangleq\sum_{x\in\mathcal{X}}A_{\hat{x}_{j},x}^{-1}\mu_{j}(x)$, then we can rewrite this as
\begin{align}
Z(\vecnot{\mu}) & =\sum_{x_{1}^{n}\in\mathcal{X}^{n}}f(x_{1},x_{2},\ldots,x_{n})\prod_{j=1}^{n}\sum_{\hat{x}_{j}\in\mathcal{X}}A_{x_{j},\hat{x}_{j}}\hat{\mu}_{j}(\hat{x}_{j})\\
 & =\sum_{\hat{x}_{1}^{n}\in\mathcal{X}^{n}}\underbrace{\left[\sum_{x_{1}^{n}\in\mathcal{X}^{n}}f(x_{1},x_{2},\ldots,x_{n})\prod_{j=1}^{n}A_{x_{j},\hat{x}_{j}}\right]}_{\triangleq\hat{f}(\hat{x}_{1},\hat{x}_{2},\ldots,\hat{x}_{n})}\prod_{j=1}^{n}\hat{\mu}_{j}(\hat{x}_{j})\\
 & =\sum_{\hat{x}_{1}^{n}\in\mathcal{X}^{n}}\hat{f}(\hat{x}_{1},\hat{x}_{2},\ldots,\hat{x}_{n})\prod_{j=1}^{n}\hat{\mu}_{j}(\hat{x}_{j})\triangleq\hat{Z}(\vecnot{\hat{\mu}}),
\end{align}
where $\hat{f}(\hat{x}_{1},\hat{x}_{2},\ldots,\hat{x}_{n})$ is the transformed factor. 
This transformation provides a change of basis for the marginalization process on a factor graph and the messages in belief propagation~\cite{AlBashabsheh-it11}.

This technique is quite useful when $\mathcal{X}$ is a finite field and $f(x_{1},\ldots,x_{n})$ is the indicator function of a subspace $S\subseteq\mathcal{X}^{n}$. In this case, the matrix $A$ is typically chosen to be the Fourier transform associated with the additive group of $\mathcal{X}$. With this choice, $\hat{f}$ is called the dual factor of $f$ and $\hat{f}(\hat{x}_{1},\hat{x}_{2},\ldots,\hat{x}_{n})$ becomes a scaled indicator function for the dual space $S^{\perp}$. 

Consider the finite field with $\left|\mathcal{X}\right|=q=p^{m}$ elements for prime $p$. It is well-known that the additive group of $\mathcal{X}$ is isomorphic to the set $\{0,1,.\ldots,p-1\}^{m}$ of vectors with elementwise modulo-$p$ addition. Thus, we assume without loss of generality that $\mathcal{X}=\{0,1,.\ldots,p-1\}^{m}$ and define the Fourier transform 
\begin{align}
A_{x,\hat{x}}=\frac{1}{\sqrt{q}}e^{-2\pi i\left\langle x,\hat{x}\right\rangle /p},
\end{align}
where $\left\langle x,\hat{x}\right\rangle _{\mathcal{X}}$ is the standard inner product between these two length-$m$ vectors. Using this convention, 
\begin{align}
\sum_{\hat{x}\in\mathcal{X}}A_{x,\hat{x}}A_{\hat{x},x'}^{-1}=\delta_{x,x'}.
\end{align}
To see the duality between indicator functions, we let the subspace $S=\left\{ \vecnot uG\,|\,\vecnot u\in\mathcal{X}^{k}\right\} $ be defined by a $k\times n$ generator matrix $G$ over $\mathcal{X}$ and we extend the inner product to $\mathcal{X}^{n}$ with $\left\langle \vecnot x,\hat{\vecnot x}\right\rangle _{\mathcal{X}^{n}}\triangleq\sum_{j=1}^{n}\left\langle x_{j},\hat{x}_{j}\right\rangle _{\mathcal{X}}$. Then, we can write 
\begin{align}
\hat{f}(\hat{x}_{1},\hat{x}_{2},\ldots,\hat{x}_{n}) & =\sum_{x_{1}^{n}\in\mathcal{X}^{n}}f(x_{1},x_{2},\ldots,x_{n})\prod_{j=1}^{n}A_{x_{j},\hat{x}_{j}}\\
 & =\sum_{x_{1}^{n}\in\mathcal{X}^{n}} \mathbb{I}(x_1^n \in S) \prod_{j=1}^{n}\frac{1}{\sqrt{q}}e^{-2\pi i\left\langle x_{j},\hat{x}_{j}\right\rangle _{\mathcal{X}}/p}\\
 & =q^{-n/2}\sum_{x_{1}^{n}\in S}e^{-\frac{2\pi i}{p}\sum_{j=1}^{n}\left\langle x_{j},\hat{x}_{j}\right\rangle _{\mathcal{X}}}\\
 & =q^{-n/2}\sum_{u_{1}^{k}\in\mathcal{X}^{k}}e^{-\frac{2\pi i}{p}\left\langle \vecnot uG,\hat{\vecnot x}\right\rangle _{\mathcal{X}^{n}}}\\
 & =\begin{cases}
q^{-n/2}q^{k} & \mbox{if }\left\langle \vecnot uG,\hat{\vecnot x}\right\rangle _{\mathcal{X}^{n}}=0\mbox{ for all }\vecnot u\in\mathcal{X}^{k}\\
0 & \mbox{otherwise.}
\end{cases}
\end{align}
The first case holds because, if $\left\langle \vecnot uG,\hat{\vecnot x}\right\rangle _{\mathcal{X}^{n}}=0$ for all $\vecnot u$, then the exponential is 1 for all $\vecnot u$ and the sum has $q^{k}$ terms. For the second case, we observe that, if there is some $\vecnot u$ such that $\left\langle \vecnot uG,\hat{\vecnot x}\right\rangle _{\mathcal{X}^{n}}\neq0$, then
\begin{align}
\left\{ u_{1}^{n}\in\mathcal{X}^{k}\,|\,\left\langle \vecnot uG,\hat{\vecnot x}\right\rangle _{\mathcal{X}^{n}}=a\right\} =a\cdot\left\{ u_{1}^{n}\in\mathcal{X}^{k}\,|\,\left\langle \vecnot uG,\hat{\vecnot x}\right\rangle _{\mathcal{X}^{n}}=1\right\} 
\end{align}
for all $a\in\mathcal{X}.$ Thus, each set has the same size (i.e., $q^{k-1}$) and we get
\begin{align}
\sum_{u_{1}^{n}\in\mathcal{X}^{k}}e^{-\frac{2\pi i}{p}\left\langle \vecnot uG,\hat{\vecnot x}\right\rangle _{\mathcal{X}^{n}}} & =\sum_{a\in\mathcal{X}}\sum_{u_{1}^{k}:\left\langle \vecnot uG,\hat{\vecnot x}\right\rangle_{\mathcal{X}^{n}} =a}e^{-\frac{2\pi i}{p}\left\langle \vecnot uG,\hat{\vecnot x}\right\rangle _{\mathcal{X}^{n}}} \\
  & =\sum_{a\in\mathcal{X}}\sum_{u_{1}^{k}:\left\langle \vecnot uG,\hat{\vecnot x}\right\rangle _{\mathcal{X}^{n}}=a}e^{-\frac{2\pi i}{p}a} \\
 & =\sum_{a\in\mathcal{X}}q^{k-1}e^{-\frac{2\pi i}{p}a}
   =0.
\end{align}
Since the dual code is defined to be the set of all vectors whose inner product with all codewords is 0, we see that $\hat{f}(\hat{x}_{1},\hat{x}_{2},\ldots,\hat{x}_{n})$ is $q^{k-n/2}$ times the indicator function for the dual code $\mathcal{C}^{\perp}$.

\section{EXIT Function Duality}
\label{sec:exit_appendix}

Here, we consider the EXIT function duality relationship for $\MCC$ and $\MCCd$~\cite[Eqn.~74]{Renes-it18} defined in our notation by
$$ H(X_i ' | Y_{\sim i}',S') = 1 - H(X_i | Y_{\sim i},S=0)_{\rho^{X_i Y,S=0}},$$
where $Y_{\sim i}$ and $Y_{\sim i}'$ are output vectors from which $Y_i$ and $Y_i'$ have been removed.
Since this is a property of the code and the not the encoder, this relationship is unaffected by applying elementary row operations to generator and parity-check matrices.
This allows us to assume that $G$ and $H$ are in a convenient systematic form.
In particular, for any $\tilde{G} \in\{0,1\}^{k\times n}$ whose row space equals $\MCC$, we can use row reduction followed by a column permutation to get a generator $G=[I_k ,\, P]$ for an equivalent code.
Since $\tilde{G}$ is full rank, the first column is always an information position and the implied column permutation maps bit 1 to bit 1.
This also allows us to assume, without loss of generality, that $i=1$.

Once $G$ is in this form, it follows that the row space of $H=[P^T , \, I_{n-k}]$ equals $\MCCd$ after the implied column permutation.
For continuity with the BEC results in Section~\ref{sec:cc_sec_duality_bec}, we will temporarily switch to that notation.
To construct $\Ag$ and $\Bg$ matrices, we can assume further that
$$ E = \begin{bmatrix} I_k & 0_{k \times (n-k)} \end{bmatrix},  \;
F = \begin{bmatrix} 0_{(n-k) \times k} & I_{n-k} \end{bmatrix} $$
because $GE^T = I_k$, $FH^T = I_{n-k}$, and $FE^T = 0_{(n-k) \times k}$.
Thus, the implied $\Ag$ and $\Bg$ matrices have the form
$$ \Ag = \begin{bmatrix} I_k & P \\ 0_{(n-k) \times k} & I_{n-k} \end{bmatrix},
\Bg = \begin{bmatrix} I_k & 0_{k \times (n-k)} \\ P^T & I_{n-k} \end{bmatrix} $$
and automatically satisfy $\Ag \Bg^T = I_n$.

Using this setup, the primal channel coding problem is defined by observing $X =[U\;S] \Ag = UG+SF$ through the PSC as $Y$ and the dual secrecy problem is defined by observing $X' =[S'\;U'] \Bg = S'E+U'H$ as $Y'$.
To remove the effect of $Y_1$, we can assume that $Y_i$ is a PSC($\theta_i)$ observation of $X_i$, where $\theta_1 = 0$ and $\theta_i = \theta$ for $i\neq 1$.
Recall that~\eqref{eq:vne_y_duality} implies that
\begin{equation} \label{eq:exit1}
H(Y_{\sim 1}|S=0)_{\rho^{Y,S=0}} =
H(Y|S=0)_{\rho^{Y,S=0}} = H(S'|Y') = H(S'|Y_{\sim 1}',X_1'),
\end{equation}
where the entropy for the dual problem is conditional on $X_1'$ because $Y_1' = X_1'$ on account that the dual channel for $X_1'$ is a BSC with error rate $p_1 = \frac{1-\cos\theta_1}{2} = 0$.

Let $\tilde{A}$ be the matrix $\Ag$ after its first row and column have been removed.
Instead of relabeling all the indices based on this change, we note that this matrix represents the original problem except that $X_1 = U_1 = 0$ and $Y_1$ is not observed (which was already achieved by $\theta_1=0$).
Defining $\tilde{B}$ in the same way, we note that the relationship $\tilde{A}\tilde{B}^T = I$ still holds.
Thus, we can apply the duality result with the $\tilde{A}$ and $\tilde{B}$ matrices to get
\begin{align} \label{eq:exit2_lhs}
H(Y_{\sim 1}|X_1,S=0)_{\rho^{Y,X_1,S=0}}
&= H(Y_{\sim 1}|X_1=0,S=0)_{\rho^{Y,X_1,S=0}} \\
&= H(\tilde{Y}|\tilde{S}=0)_{\rho^{\tilde{Y},\tilde{S}=0}} \\
&= H(\tilde{S}'|\tilde{Y}') \\
&= H(S_{\sim 1}'|Y_{\sim 1}',S_1'=0), \label{eq:exit2_rhs}
\end{align}
where the construction of $\tilde{B}$ implies the following connections $S_{\sim 1}' \equiv \hat{S}'$, $Y_{\sim 1}' \equiv \tilde{Y}'$, and $S_1' = 0$.
If we let $Q$ be the result of subtracting the LHS of~\eqref{eq:exit2_lhs} from the LHS of~\eqref{eq:exit1}, then we find that
\begin{align}
Q &= H(Y_{\sim 1}|S=0)_{\rho^{Y,X_1,S=0}} - H(Y_{\sim 1}|X_1,S=0)_{\rho^{Y,X_1,S=0}} \\
&= I(X_1;Y_{\sim 1}|S=0)_{\rho^{Y,X_1,S=0}} \\
&= 1 - H(X_1|Y_{\sim 1},S=0)_{\rho^{Y,X_1,S=0}}
\end{align}
equals one of the entropies associated with EXIT function duality.
Of course, $Q$ also equals the result of subtracting the RHS of~\eqref{eq:exit2_rhs} from the RHS of~\eqref{eq:exit1}.
This implies that
\begin{align*}
Q &=  H(S'|Y_{\sim 1}',X_1') - H(S_{\sim 1}'|Y_{\sim 1}',S_1' = 0) \\
&= H(S'|Y_{\sim 1}',X_1') - H(S_{\sim 1}'|Y_{\sim 1}',S_1') \\
&= H(S_1 '|Y_{\sim 1}',X_1') + H(S_{\sim 1} '|Y_{\sim 1}',S_1',X_1') - H(S_{\sim 1}'|Y_{\sim 1}',S_1') \\
&= H(S_1 '|Y_{\sim 1}',X_1') - I(X_1';S_{\sim 1} '|Y_{\sim 1}',S_1') \\
&= H(S_1 '|Y_{\sim 1}',X_1') - H(X_1 ' | Y_{\sim 1}',S_1') + H(X_1'|Y_{\sim 1}',S_{\sim 1}',S_1') \\
&=  H(X_1'|Y_{\sim 1}',S'),
\end{align*}
where the last step follows from the fact that $H(S_1 '|Y_{\sim 1}',X_1') = H(X_1 ' | Y_{\sim 1}',S_1')$ because $S_1' + X_1' = [U' P^T]_1$ depends only on $U'$.  
This shows that $Q$ also equals the other entropy associated with EXIT function duality.
Thus, we can conclude that
\begin{equation}
H(X_1'|Y_{\sim 1}',S') = 1 - H(X_1|Y_{\sim 1},S=0)_{\rho^{Y,X_1,S=0}}.
\end{equation}


\end{document}